\documentclass[sigconf]{acmart}
\usepackage{tikz}
\usepackage{hyperref}
\usepackage{color,soul}

\usepackage{amsmath,amssymb,amsfonts}
\usepackage{graphicx}
\usepackage{textcomp}
\usepackage{xcolor}
\usepackage{xspace}
\usepackage{lipsum}
\usepackage{multirow}
\usepackage{booktabs}

\newcommand{\name}{\textit{TA}\xspace}

\usepackage[noend]{algorithmic}
\usepackage[ruled,vlined,linesnumbered,algo2e]{algorithm2e}

\newcommand{\Fig}[1]{Fig.~\ref{#1}}

\newcommand{\Tbl}[1]{Tbl.~\ref{#1}}
\newcommand{\Sec}[1]{Sec.~\ref{#1}}
\newcommand{\Reb}[1]{\textcolor{red}}
\newcommand{\Alg}[1]{Alg.~\ref{#1}}

\definecolor{mycolorblue}{RGB}{81,140,230}
\definecolor{mycolorgreen}{RGB}{166,213,95}
\definecolor{mycoloryellow}{RGB}{242,210,98}
\definecolor{mycolorpink}{RGB}{242,96,119}
\definecolor{mycolorpurple}{RGB}{149,90,189}
\definecolor{mycolororange}{RGB}{255,199,7}

\newcommand\circlednumberblack[1]{%
  \begin{tikzpicture}[baseline=(char.base)]
    \node[shape=circle,,fill=black,inner sep=1pt] (char) {\textcolor{white}{\scriptsize\sffamily\bfseries#1}};
\end{tikzpicture}}

\newcommand{\revise}[1]{#1}
\newcommand{\revision}[2]{#2} 
\newcommand{\recaption}[2]{#2}

\AtBeginDocument{%
  }
    
\copyrightyear{2025}
\acmYear{2025}
\setcopyright{cc}
\setcctype{by}
\acmConference[ISCA '25]{Proceedings of the 52nd Annual International Symposium on Computer Architecture}{June 21--25, 2025}{Tokyo, Japan}
\acmBooktitle{Proceedings of the 52nd Annual International Symposium on Computer Architecture (ISCA '25), June 21--25, 2025, Tokyo, Japan}\acmDOI{10.1145/3695053.3731043}
\acmISBN{979-8-4007-1261-6/2025/06}

\begin{document}

\title[Transitive Array: An Efficient GEMM Accelerator with Result Reuse]{Transitive Array: An Efficient GEMM Accelerator\\ with Result Reuse}



\author{Cong Guo}
\email{cong.guo@duke.edu}
\authornote{Cong Guo is the corresponding author of this paper.}
\affiliation{%
  \institution{Duke University}
  \city{Durham}
  \country{USA}}

\author{Chiyue Wei}
\email{chiyue.wei@duke.edu}
\affiliation{%
  \institution{Duke University}
  \city{Durham}
  \country{USA}}

\author{Jiaming Tang}
\email{jmtang@mit.edu}
\affiliation{%
  \institution{MIT}
  \city{Cambridge}
  \country{USA}}

\author{Bowen Duan}
\email{bowen.duan@duke.edu}
\affiliation{%
  \institution{Duke University}
  \city{Durham}
  \country{USA}}

\author{Song Han}
\email{songhan@mit.edu}
\affiliation{%
  \institution{MIT}
  \city{Cambridge}
  \country{USA}}

\author{Hai Li}
\email{hai.li@duke.edu}
\affiliation{%
\institution{Duke University}
\city{Durham}
\country{USA}}

\author{Yiran Chen}
\email{yiran.chen@duke.edu}
\affiliation{%
\institution{Duke University}
\city{Durham}
\country{USA}}




\begin{abstract}
Deep Neural Networks (DNNs) and Large Language Models (LLMs) have revolutionized artificial intelligence, yet their deployment faces significant memory and computational challenges, especially in resource-constrained environments. 
Quantization techniques have mitigated some of these issues by reducing data precision, primarily focusing on General Matrix Multiplication (GEMM). 
This study introduces a novel sparsity paradigm, \textbf{transitive sparsity}, which leverages the reuse of previously computed results to substantially minimize computational overhead in GEMM operations. 
By representing transitive relations using a directed acyclic graph, we develop an efficient strategy for determining optimal execution orders, thereby overcoming inherent challenges related to execution dependencies and parallelism.
Building on this foundation, we present the \textbf{Transitive Array}, a multiplication-free accelerator designed to exploit transitive sparsity in GEMM. 
Our architecture effectively balances computational workloads across multiple parallel lanes, ensuring high efficiency and optimal resource utilization. 
Comprehensive evaluations demonstrate that the Transitive Array achieves approximately \revise{\textbf{7.46$\times$}} and \revise{\textbf{3.97$\times$}}  speedup and \revise{\textbf{2.31$\times$}} and \revise{\textbf{1.65$\times$}} energy reduction compared to state-of-the-art accelerators such as Olive and BitVert while maintaining comparable model accuracy on LLaMA models.

\end{abstract}



\maketitle

\section{Introduction}
\label{sec:intro}
Deep neural networks (DNNs)\cite{lecun2015deep, goodfellow2016deep} have become foundational to modern artificial intelligence, demonstrating transformative potential across various applications. 
Among these, large language models (LLMs)\cite{devlin2018bert, brown2020languagemodelsfewshotlearners} stand out for their ability to generate human-like text, understand complex language, and perform advanced reasoning, making them crucial in natural language processing\cite{vaswani2017attention, brown2020language}.
However, as DNNs scale, they introduce substantial memory and computational overheads that challenge their deployment efficiency, particularly in resource-constrained environments~\cite{shoeybi2020megatronlmtrainingmultibillionparameter, kaplan2020scalinglawsneurallanguage}.

Quantization~\cite{jacob2017quantizationtrainingneuralnetworks} has thus emerged as a critical research avenue, offering methods to reduce model size and computational load by representing parameters with lower precision while aiming to retain model accuracy. 
Most quantization approaches aim to reduce the data precision of general matrix multiplication (GEMM) operations, which are fundamental to DNNs~\cite{chetlur2014cudnnefficientprimitivesdeep, jouppi2017indatacenterperformanceanalysistensor}, as they represent the core computational workload in both training and inference. 
Recently, quantization has regained popularity in LLMs. 
AWQ~\cite{lin2024awqactivationawareweightquantization} achieves a 4-bit compression of weights with negligible accuracy loss. 
Additional advanced algorithms (e.g., BitNet1.58~\cite{wang2023bitnetscaling1bittransformers}) push the limits further by targeting 1-bit weight quantization. 
Through such innovations, quantization significantly enhances LLM efficiency and deployability.

\begin{figure}[t]
    \centering
    \includegraphics[width=0.98\linewidth]{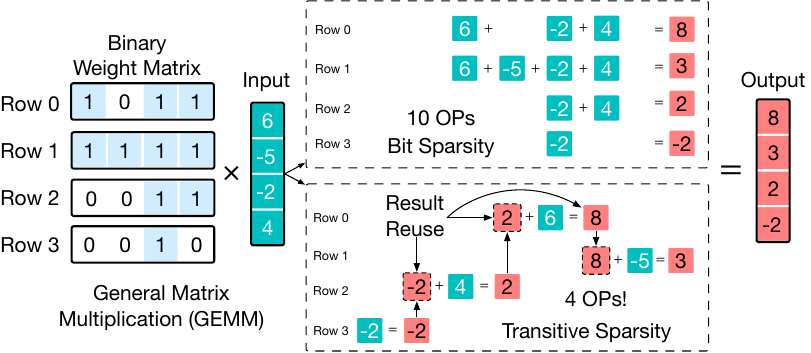}   
    \vspace{-3mm} 
    \caption{\protect\revision{CQ1a}{Comparison of bit sparsity and transitive sparsity (Ours) on the binary general matrix multiplication.}}
    \label{fig:intro}
    \vspace{-6mm}
\end{figure}

Building on quantization, this study explores novel opportunities to accelerate GEMM operations.
\recaption{CQ1a}{Bit-slicing has proven to be an effective technique in DNN accelerators~\cite{bitlet,bitwave,chen2024bbsbidirectionalbitlevelsparsity,pragmatic}. 
It decomposes each bit of a quantized matrix into multiple binary 1-bit matrices, as illustrated in \Fig{fig:bitslice}. 
By leveraging bit-slicing, we transform quantized {integer-based} GEMM into binary GEMM, as shown in \Fig{fig:intro}.}
Bit-slice accelerators exploit bit-level sparsity, achieving computational reductions of up to 50–60\%~\cite{chen2024bbsbidirectionalbitlevelsparsity}. 
However, our analysis reveals even greater sparsity potential beyond bit sparsity, highlighting further opportunities for optimization.

This study introduces a novel sparsity paradigm termed \textbf{transitive sparsity}, which leverages previously computed results to minimize overall computations. 
As illustrated in \Fig{fig:intro}, consider Row-0 of the \revise{binary} matrix, represented by the bit pattern $1011$, which requires the accumulation of values $6$, $-2$, and $4$, and Row-2, represented by $0011$, which involves the accumulation of $-2$ and $4$. 
Since both rows share the accumulation of $-2$ and $4$, we can reuse the computed result of Row-2 for Row-0, significantly reducing the accumulation operations in GEMM. 
In the example shown in \Fig{fig:intro}, transitive sparsity enables a 2.5-fold reduction in operations (from 10 to 4) compared to bit sparsity and a 4-fold decrease over dense GEMM.
Furthermore, it is worth noting that transitive sparsity is algorithm-agnostic and lossless,  preserving the computational accuracy of quantized GEMM operations.
Our evaluation demonstrates that transitive sparsity theoretically reduces overall computations by $8\times$ (i.e., $87.5\%$ sparsity) for the LLaMA-7B~\cite{touvron2023llama1} compared to dense GEMM without accuracy loss after the quantization.

To harness the benefits of transitive sparsity, we propose a novel computer architecture called the \textbf{transitive array}. 
However, we encountered several challenges in efficiently implementing transitive sparsity.
First, transitive sparsity imposes strict requirements on execution order. For example, if Row-0 is computed before Row-2, it prevents reuse of Row-2’s results. 
For weight tensors, we can precompute an optimal execution order offline. 
However, attention operations~\cite{vaswani2023attentionneed} in LLM present a complication, as they involve dynamically generated activations (e.g., the Query and Key tensors). 
This dynamic nature makes it challenging to determine execution order on the fly without incurring significant overhead.
This dynamic limitation is also observed in most existing accelerators~\cite{chen2024bbsbidirectionalbitlevelsparsity, guo2023olive, bitlet}, which consequently lack adequate support for attention layers in their designs. 
Instead, these accelerators predominantly focus on fully connected (FC) layers with pre-processed weights.
Second, even with an optimized execution order, the strict dependencies within transitive sparsity result in inherently serial execution, greatly restricting parallelism. 
Additionally, the irregularity of sparsity patterns makes it difficult to balance workloads across multiple parallel lanes, as varying sparsity levels lead to inconsistent computational loads and underutilization.

Fortunately, we identified certain properties within transitive sparsity that help address these challenges. 
Leveraging these insights, we can represent the transitive relations of the transitive sparsity using a directed acyclic graph, precisely a Hasse graph~\cite{hasse_diagram_wikipedia, hasse1967grundlagen}. 
This representation enables us to reduce the overhead of generating an optimal execution order, achieving linear complexity compared to the cubic complexity typically associated with GEMM.
Following this, we assign rows of the bit-sliced matrix across multiple parallel lanes and have proven that this approach eliminates dependencies among parallel lanes. 
Additionally, to address workload imbalance, we introduce a simple yet effective method to ensure balanced distribution across multiple lanes before execution, making sparse execution more regular and highly efficient.

This work presents a comprehensive solution and novel architecture to enable transitive sparsity in GEMM operations. 
This work provides a new perspective on transitive sparsity and insights for future designs of more versatile GEMM accelerators. 
Compared to state-of-the-art accelerators like Olive~\cite{guo2023olive}, which leverages outlier-aware quantization, and BitVert~\cite{chen2024bbs}, which employs bit-slice techniques, our solution achieves an approximate \revise{\textbf{7.46$\times$}} and \revise{\textbf{3.97$\times$}} speedup and \revise{\textbf{2.31$\times$}}, \revise{\textbf{1.65$\times$}} energy reduction, under similar model accuracy of LLaMA~\cite{touvron2023llama1}.
Our contributions are summarized as follows:
\begin{itemize}
    \item We define a novel sparsity paradigm, \textbf{Transitive Sparsity}, which leverages previously computed results to minimize GEMM computations, thereby achieving significant reductions in operational overhead.
    \item We design a novel computer architecture, \textbf{Transitive Array (TA)}, tailored to exploit transitive sparsity and effectively address challenges related to execution order and parallelism. Our design also efficiently supports the quantization of Attention layers.
    \item Transitive Array is \textbf{multiplication-free}, achieving high efficiency by eliminating multiplication operations.
    \item Through comprehensive evaluations, we demonstrate the effectiveness of the \textbf{Transitive Array} architecture, showcasing substantial speedups and energy reductions compared to existing accelerators.
\end{itemize}
\section{Background and Motivation}
This section presents the motivation for transitive sparsity, grounded in our novel observations of sparsity patterns within GEMM. 

\subsection{\revise{Background: Quantization and Bit-slicing}}
\label{sec:prelim}

\begin{figure}[t]
    \vspace{-2mm}
    \centering
    \includegraphics[width=1\linewidth]{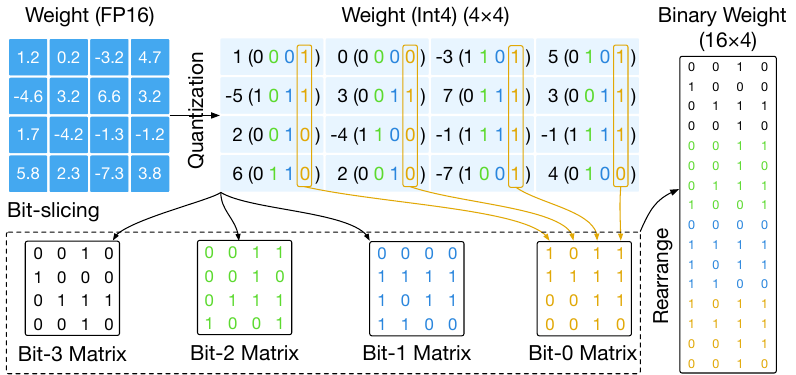}
    \vspace{-6mm}
    \caption{\recaption{CQ1b}{Quantization and bit-slicing. A weight tensor (FP16) is quantized into an Int4 weight matrix ($4 \times 4$). Bit-slicing decomposes each bit level of the Int4 matrix and reorganizes them into a binary weight matrix of shape ($16 \times 4$).}}   
    \label{fig:bitslice}
    \vspace{-5mm}
\end{figure}

\revision{CQ1b}{
As illustrated in \Fig{fig:bitslice}, quantization and bit-slicing~\cite{bitlet, bitwave, pragmatic, chen2024bbsbidirectionalbitlevelsparsity} serve as foundational techniques for implementing transitive sparsity. After quantization, a 4-bit integer matrix of shape ($4 \times 4$) is decomposed into four bit-level matrices. These matrices are then reorganized into a binary weight matrix with a shape of ($16 \times 4$).
}

\revise{
In general, for an $S$-bit quantized matrix of shape ($N \times K$), we can reorganize it into a ($S\cdot N \times K$) binary matrix.}
Building on previous work~\cite{pragmatic, bitwave}, we demonstrate that maintaining sufficient precision in addition and accumulation operations—thereby preventing overflow or underflow—ensures that \revise{the bit-slicing method achieves the same results as the original quantized GEMM without any loss, even when adjusting the accumulation/reduction order for integers.}

\subsection{\revise{Motivation: Transitive Sparsity}}
\label{sec:motivation}
\begin{figure}[t]
    \vspace{-2mm}
    \centering
    \includegraphics[width=0.85\linewidth]{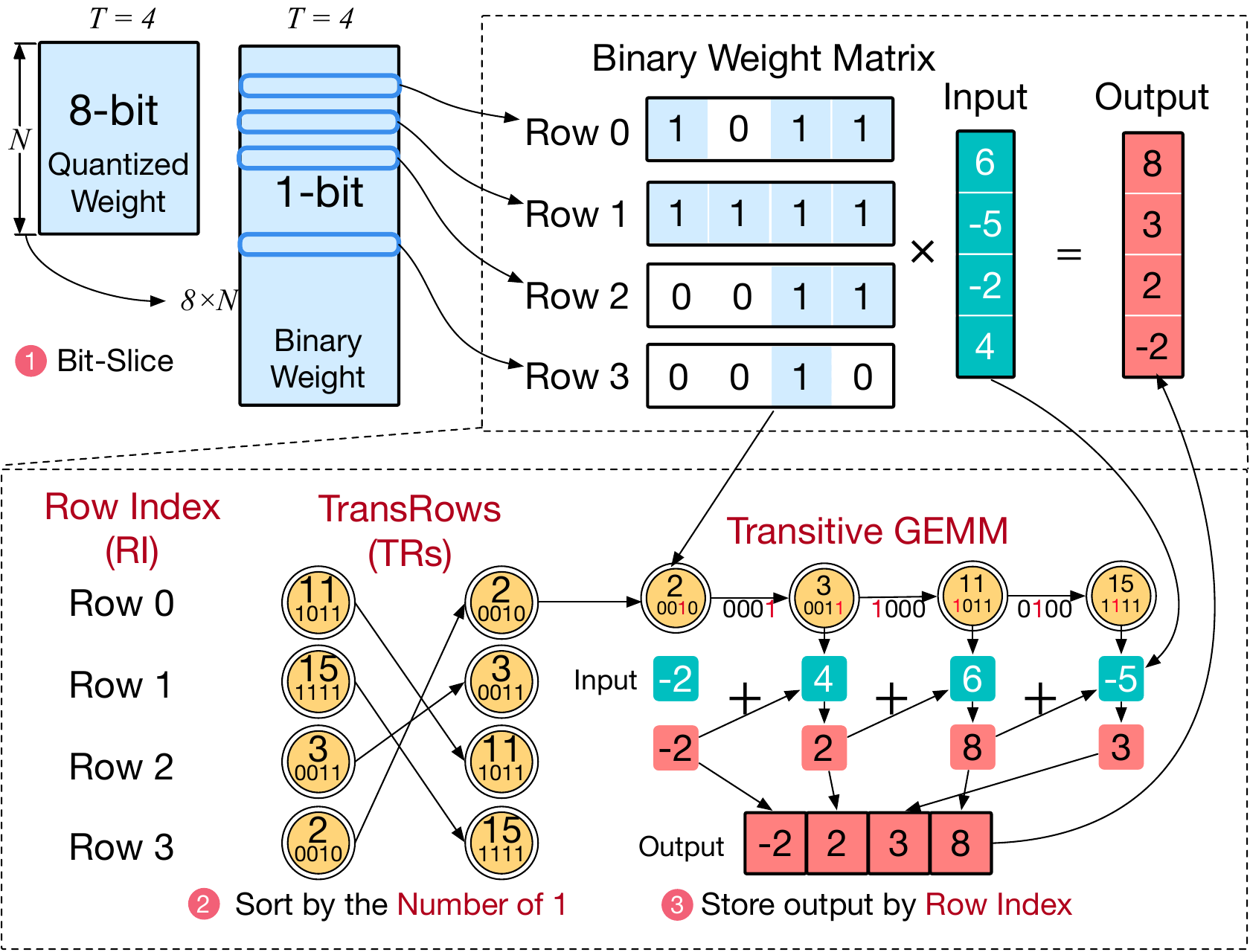}
    \vspace{-4mm}
    \caption{\recaption{CQ1b}{Motivation of Transitive GEMM.}}
    \label{fig:motivation}
    \vspace{-6mm}
\end{figure}

\recaption{CQ1b}{ }
Following these preliminaries, \Fig{fig:motivation} illustrates our motivation for transitive sparsity.
\revise{We first slice (\circlednumberblack{1}) an 8-bit weight tensor with $N$ rows and $T = 4$ columns into a matrix with $8\times N$ rows and $T$ columns, where each row represents a binary row with the $T$-bit width.
We define each binary row as a \textbf{Transitive Row (TransRow, TR)}, which serves as the fundamental unit of our design.
The width of each TransRow denoted as $T$, plays a crucial role in determining the efficiency of our approach.
} 

In this study, we exemplify transitive sparsity using a 4-bit TransRow width. 
Transitive sparsity is also prevalent in higher bit widths (e.g., 8-bit).
Each TransRow is identified by an index that indicates its corresponding bit level in the original integer. 
These TransRows are then accumulated with appropriate shifts. 
We employ 2's complement representation for integers.
In practice, we represent all one-bits as positive 1, and all TransRows can be represented by \revise{\textbf{unsigned integers}, as illustrated in \Fig{fig:motivation}~\circlednumberblack{2}}.

Subsequently, specific rows can be selected. 
For instance, consider four TransRows with values \revise{11 (\texttt{1011}), 15 (\texttt{1111}), 3 (\texttt{0011}), and 2 (\texttt{0010})}.
Notably, some TransRows share identical 1s. 
For example, the TransRow representing 11 (\texttt{1011}) encompasses the same 1s as the TransRow representing 3 (\texttt{0011}). 
Exploiting this property allows us to eliminate redundant computations for TransRows. 
Originally, the TransRow 11 (\texttt{1011}) required three accumulations of $6$, $-2$, and $4$, skipping the 0 bit. 
\revise{As illustrated in \Fig{fig:motivation}~\circlednumberblack{3},} by reusing the result, we only need to perform an additional accumulation for $2$ (the result of TransRow \texttt{0011}) and $6$, corresponding to the difference bits \texttt{1000} (i.e., \revise{\texttt{1011}$\oplus$ \texttt{0011} $=$ \texttt{1000}}). 
We term this approach \textbf{Transitive Sparsity}, which reuses the results of preceding TransRows. 
Consequently, this characteristic of the bit-weight tensor significantly reduces the computational load for GEMM.

\revise{\textbf{Challenges.}}
However, as discussed in \Sec{sec:intro}, adopting transitive sparsity in Deep Neural Networks (DNNs) introduces several challenges. 
For example, as shown in \Fig{fig:motivation}(b), the execution order of the four TransRows deviates from the original sequence. 
This strict execution order requirement renders parallel computing infeasible. 
Additionally, generating the execution order is highly complex.
Specifically, we must determine which pairs of TransRows satisfy the transitive sparsity criteria among potentially hundreds of rows, leading to significant overhead in computing the execution order. 
A naive implementation would result in substantial inefficiencies in hardware utilization or increased time overhead. In this study, we aim to address these issues efficiently.

\subsection{Efficient Representation}

\begin{figure}[t]
    \centering
    \includegraphics[width=1\linewidth]{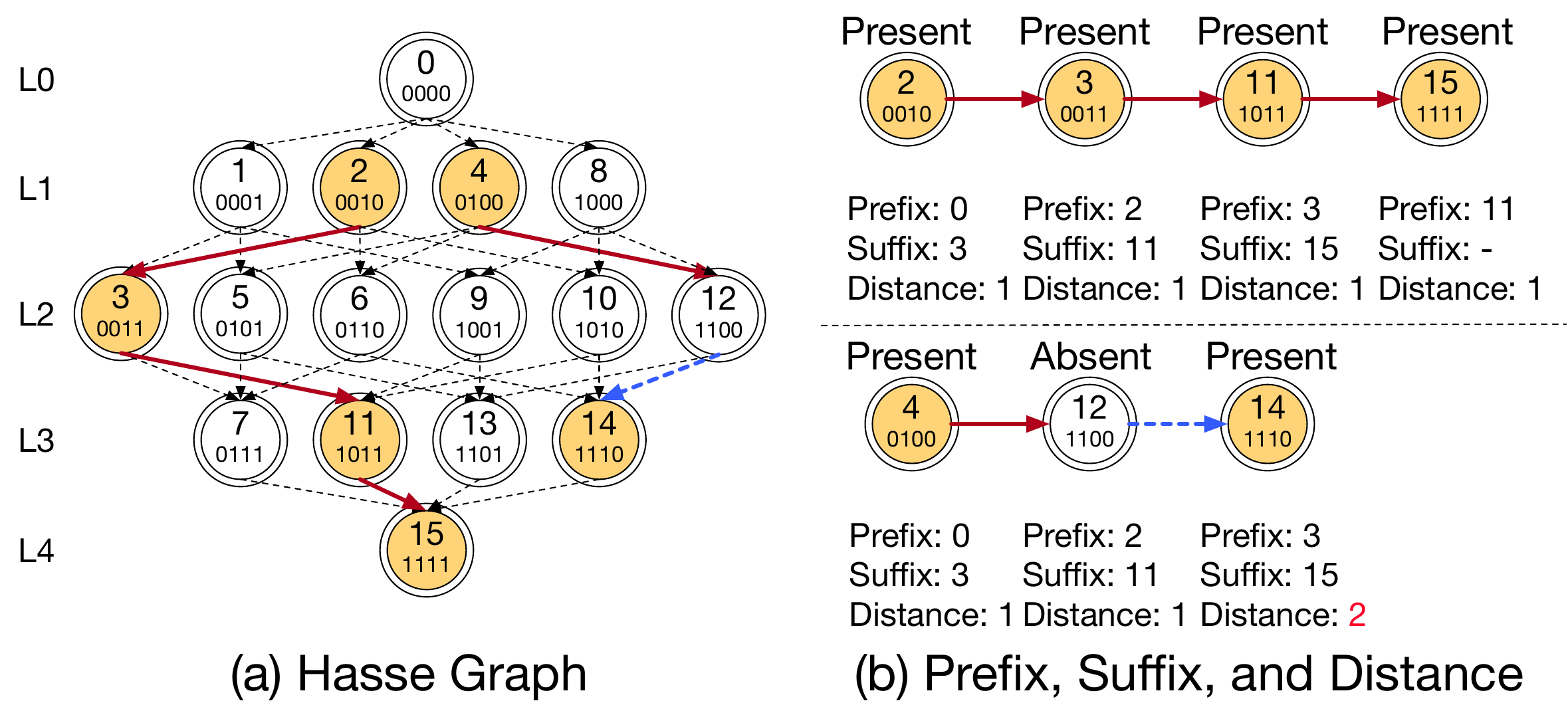}
    \vspace{-6mm}
    \caption{\recaption{CQ1c}{Hasse graph and definition of prefix, suffix, and distance.}}
    \label{fig:hasse}
    \vspace{-5mm}
\end{figure}

Fortunately, we have identified a beneficial and intriguing property within GEMM. 
As illustrated in \Fig{fig:motivation}, neighboring TransRows inherently possess a partial ordering relationship characterized by a single bit flip. 
This indicates that result reuse exhibits strong partial order relations.
Inspired by this observation, we represent transitive sparsity using a \textbf{Hasse graph}~\cite{hasse1967grundlagen}, a specialized \textbf{Directed Acyclic Graph} that depicts a finite partially ordered set. 
\Fig{fig:hasse}(a) illustrates all partial order relations within a 4-bit TransRow. 
Normally, we ignore the Level $0$, which can be skipped without computation.
This representation effectively and significantly reduces the complexity of our design, which will be detailed in the following section.

To conveniently present our design, we introduce three concepts: \textbf{prefix}, \textbf{suffix}, and \textbf{distance} within the Hasse graph, as shown in \Fig{fig:hasse}(b).
We define the preceding node (e.g., 3) that provides the reused result as the \textbf{prefix}, correspondingly, the latter as the \textbf{suffix}, e.g., Node 11. 
Based on transitivity, a further subsequent Node 15 can continue to reuse the result of its prefix, i.e., 11.

The \textbf{distance} between two nodes with a partial ordering is defined as the difference in their levels, corresponding to the difference in the number of 1s they contain. 
As illustrated in \Fig{fig:hasse}(b)-bottom, the \textbf{distance} is transitive and depends on the presence of specific TransRows and nodes. 
The distance is adjusted if an intermediate node is absent from the TransRows. 
For example, the distance between \revise{Node 4 and Node 14 is 2 because Node 12 is absent. 
However, Node 14 still considers Node 12 as its prefix, requiring Node 12 to pass the result from Node 4 to Node 14.} 
Therefore, a larger distance implies that more add operations are required from the corresponding prefix TransRow. 
When the distance is 1, we achieve the most efficient result reuse between two different TransRows. 
Additionally, TransRows with identical values have a distance of 0 and can directly reuse each other's results.

\subsection{Parallelism and Load Balance}
After effectively representing transitive relations using a Hasse graph, we observe that nodes within the same level possess no partial ordering relationships, as they cannot share results. 
For instance, Node 2 cannot share its result with Node 4 because they differ in their bit representations. 
Consequently, transitive sparsity exhibits parallelism horizontally across each level of the Hasse graph.

\textbf{Parallelism.}
Specifically, for an $S$-bit width transitive sparsity, the maximum parallelism is achieved at Level-$\frac{S}{2}$, corresponding to the binomial coefficient \( \binom{S}{\frac{S}{2}} = \frac{S!}{\left(\frac{S}{2}\right)! \cdot \left(\frac{S}{2}\right)!} \). 
For example, Level $2$ in a 4-bit graph exhibits a parallelism of 6, while Level $4$ in an 8-bit graph exhibits a parallelism of 70. 
However, leveraging the maximum parallelism would result in the underutilization of other levels. 
Therefore, we typically utilize the granularity corresponding to Level $1$, which is 4 for a 4-bit width and 8 for an 8-bit width TranSparsity. 
This strategic choice of granularity ensures balanced utilization across all levels of the Hasse graph, optimizing both computational efficiency and hardware resource allocation.

\textbf{Data Independence.}
Additionally, each node in the Hasse graph can be transitioned by only one prefix node. 
For example, Node 11 can be transitioned from only one of its prefix nodes, such as Node 3, Node 9, or Node 10. 
Therefore, we can assign only one prefix to every node in the graph. 
Consequently, we can divide the graph into a forest comprising $T$ independent trees starting from Level $1$ for $T$-bit transitive sparsity.
Therefore, we can safely divide the Hasse graph into multiple independent trees, each node having one in-degree edge, as shown in \Fig{fig:hasse}(a). 
This division ensures parallelism and that each node has a prefix node with correct partial ordering.

\textbf{Load Balance.}
After dividing the Hasse graph to enable parallelism, it is essential to balance the resulting trees to prevent extreme workload imbalances across parallel lanes, which can lead to inefficient resource utilization and performance bottlenecks. 
We design an effective round-robin-like traversal method by leveraging the rule that each node has only one prefix. 
This method allows us to select an available prefix node for each node, thereby evenly distributing workloads among the trees only using a simple node number counter for supervision. 

Finally, this section presents a streamlined representation of transitive sparsity and theoretically addresses the challenges of parallelism and load balancing. 
This framework paves the way for more efficient architectural designs, which are detailed in the subsequent sections.

\begin{figure*}[t]
    \centering
    \includegraphics[width=1\linewidth]{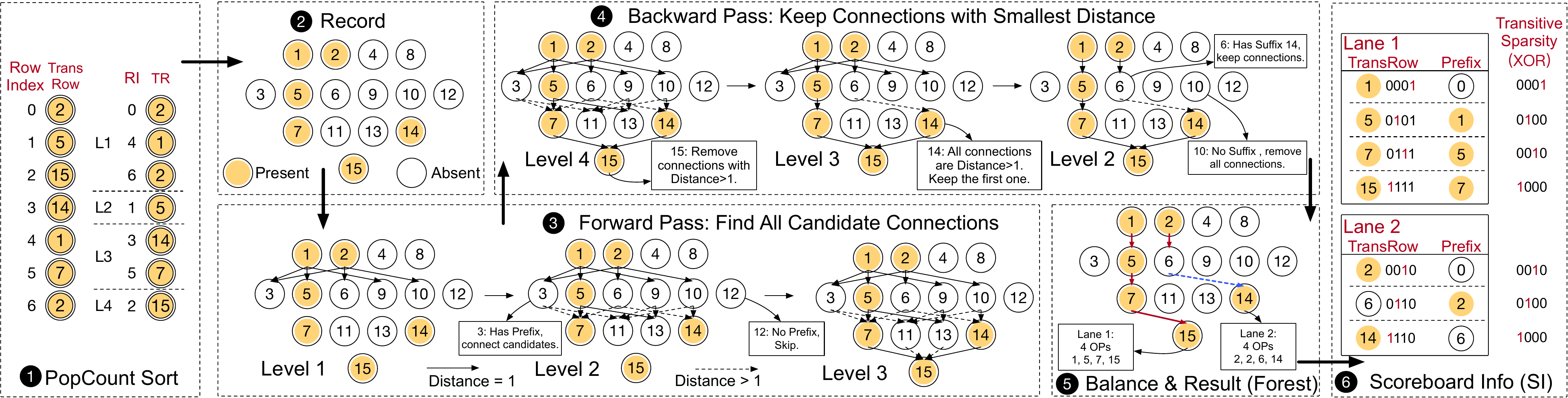}
    \vspace{-6mm}
    \caption{\protect\revision{CQ1c}{Scoreboarding process.
Step~\protect\circlednumberblack{1}: Sort TransRows by Hamming order.
Step~\protect\circlednumberblack{2}: Record present TransRows in the Hasse graph.
Step~\protect\circlednumberblack{3} (Forward pass): Assign candidate prefixes to all nodes.
Step~\protect\circlednumberblack{4} (Backward pass): Propagate suffix requests for nodes without a Distance $=1$ prefix, searching for the prefix with the shortest distance and retaining the closest prefix for all nodes.
Step~\protect\circlednumberblack{5}: Generate a balanced forest.
Step~\protect\circlednumberblack{6}: Output the final Scoreboard information (SI).}
}

\vspace{-4mm}
    \label{fig:scoreboard_process}
    \Description{}
\end{figure*} 
\section{Scoreboard Design}
\label{sec:scoreboard}
In this section, we introduce the \textbf{Scoreboard} mechanism, designed to efficiently compute the execution order using a Hasse graph representation.
\recaption{CQ2b}{
We propose two types of Scoreboard: \textbf{Static} and \textbf{Dynamic}.
The static Scoreboard operates offline, generating execution information at the tensor level.
In contrast, the dynamic Scoreboard processes smaller sub-GEMM operations online, ensuring compatibility with all GEMM kernels and enabling seamless deployment for large language models (LLMs).}
To accommodate the dynamic nature of attention mechanisms in Transformer architectures, the Scoreboard efficiently determines the execution order through dedicated hardware design, while simultaneously optimizing parallelism and workload balance at runtime.
\revise{Both static and dynamic mechanisms leverage the same Scoreboard algorithm, as illustrated in \Fig{fig:scoreboard_process}, \Alg{alg:forward}, and \Alg{alg:backward}}

\subsection{Hamming-order Execution}
Unlike general applications, which handle irregular and dynamic instructions, the instructions and data accesses in GEMM operations within DNN models exhibit strict regularity. 
This regularity enables us to predefine execution orders for all instructions before they are issued.
Based on our analysis using the Hasse graph, the level of a node corresponds to the number of 1s (i.e., PopCount) in its binary representation. 
For instance, Node 11 (\texttt{1011}) belongs to Level 3.
Leveraging this property, we sort the incoming \textit{TransRows} based on their Hamming weight (i.e., PopCount)~\cite{knuth2009art} instead of their integer values, defining this execution order as \textbf{Hamming-order}.

As shown in \Fig{fig:scoreboard_process}~\circlednumberblack{1}, TransRows are sorted exclusively by their \textbf{\revise{number of 1s}}. 
Moreover, since there is no inherent partial ordering within nodes at the same level, the sorting mechanism does not enforce any order among nodes with identical PopCount. 
This approach significantly simplifies the sorter design, reducing its complexity and hardware overhead.

\subsection{Scoreboarding}
\label{sec:scoreboarding}
To achieve a balanced forest with independent trees, we design an efficient Scoreboarding algorithm optimized for hardware implementation, as depicted in \Fig{fig:scoreboard_process}~\circlednumberblack{2}-\circlednumberblack{5}. 

First, we record (Step~\circlednumberblack{2}) \revise{all present TransRow and their counts in the Hasse graph}. 
Subsequently, we perform \textbf{forward} and \textbf{backward} passes to build and prune the Hasse graph for transitive sparsity, maintaining prefix information with the smallest distance for each node.

\textbf{Forward Pass for Prefix.}
As shown in \Alg{alg:forward} and Step~\circlednumberblack{3}, we \revise{hierarchically} update each node's prefix information without interdependencies.
\revision{RD7}{Therefore, we need to traverse all nodes by the Hamming order (Line 3) in Step~\circlednumberblack{3}.}
Initially, we initialize the nodes (Lines~1-2) and traverse all nodes in a forward partial ordering (Lines~3-4). 
For transitive sparsity, we select node prefixes with a distance of less than 4 (Line~7). 
In practice, even in an 8-bit TransRow, nodes with a distance of 3 are rarely observed (\revision{CQ1e}{$<0.1\%$ within 128 TransRows}).
In Line~8, we set a temporary distance of 0 for nodes with a valid distance and \textit{Count} $> 0$. 
This indicates that the node will be executed and has a valid prefix, as at least one TransRow has been processed. 
Consequently, the node can serve as a prefix, resetting the distance for subsequent suffixes. 
Following this, we \revise{generate locally the suffix based on the Hasse graph} and send the prefix information to all its suffixes with an incremented \revise{($+1$)} distance (Lines~9-10), embodying the transitivity of TranSparsity. 
Lines~11-13 set the distance and prefix bitmaps.

For example, as depicted in \Fig{fig:scoreboard_process}~\circlednumberblack{3}, Nodes 1 and 2 will have their temporary distances set to 0 due to their \textit{Count} > 0. 
Subsequently, Node 5 receives prefix information from Node 1 with a distance of 1, recording them into prefix bitmap. 
Similarly, all nodes update their prefix information with corresponding distances. 
A special case is Node 14, which has a distance of 2 from Node 2. Due to its \textit{Count} $> 0$, Node 14 still sends a distance of 1 to Node 15. This mechanism retains more available prefix information, forming a more effective graph.

\begin{algorithm2e}[t]
    \caption{4-bit Scoreboarding Forward Pass.}
    \label{alg:forward}
    \small
\KwIn{Node list, $N[16]$
\{
    int Count;
    int Distance;
    Bitmap Prefix[4];
    Bitmap Suffix;
\}
}


\SetKwFunction{Forward}{Forward}
\SetKwFunction{SetPrefix}{SetPrefix}

\SetKwProg{Fn}{def}{:}{}
\For {i in [\revise{1, 2, 3, 4, 5, 6, 7, 8, 9, 10, 11, 12, 13, 14, 15}]}
{
    N[i].Distance = $+\infty$\\
}
\For {i in [0, 1, 2, 4, 8, 3, 5, 6, 9, 10, 12, 7, 11, 13, 14]}
{
    \Forward{i}
}
\Fn{\Forward{int idx}}{
    int dis = N[idx].Distance \\
    \textbf{if} (dis >= 4 and idx != 0) \textbf{then} return\\
    \textbf{if} ((N[idx].Count > 0) or idx == 0) \textbf{then} dis = 0\\
    \While{suffix $\in$ \revise{GetSuffixFromHasse}(idx)}{
        \SetPrefix{suffix, dis + 1, idx}
    }
}
\Fn{\SetPrefix{int idx, int Distance, int prefix}}{
    \revise{N[idx].Prefix[Distance - 1].insert(prefix)} \\
    N[idx].Distance = min(N[idx].Distance, Distance)\\
}

\end{algorithm2e}

\begin{algorithm2e}[t]
    \caption{4-bit Scoreboarding Backward Pass.}
    \label{alg:backward}
    \small
\KwIn{Node list, $N[16]$
\{
    int Count;
    int Distance;
    Bitmap Prefix[4];
    Bitmap Suffix;
\}
}


\SetKwFunction{Backward}{Backward}
\SetKwFunction{SetSuffix}{SetSuffix}

\SetKwProg{Fn}{def}{:}{}
\For {i in [15, 7, 11, 13, 14, 3, 5, 6, 9, 10, 12, 1, 2, 4, 8]}
{
    \Backward{i}
}

\Fn{\Backward{int idx}}{
    int dis = N[idx].Distance\\
    \If{1 < dis < 4 and N[idx].Count > 0}{
        prefix = \revise{GetPrefixFromHasse}(N[idx].Prefix[dis - 1])\\
        \SetSuffix{prefix[0], idx} \tcp*[h]{Only the first prefix.}\\
    }    
}
\Fn{\SetSuffix{int idx, int suffix}}{
    \revise{N[idx].Suffix.insert(suffix)}\\
    N[idx].Count = 1
}
Keep PB with the smallest distance; remove others.

\end{algorithm2e}

\textbf{Backward Pass for Suffix.}
As shown in \Alg{alg:backward} and Step~\circlednumberblack{4}, the backward pass is a reverse process of the forward pass to retrieve suffix information. 
The algorithm primarily constructs paths for nodes with \textit{Count} $> 0$ and distance $>1$ to their valid prefix nodes (Lines~5-7). 
Notably, we select only one prefix (defaulting to the first available) from the first valid Prefix Bitmap (PB) field to build the path (Line~7). Selecting multiple prefixes could result in redundant paths for a node.
For instance, in \Fig{fig:scoreboard_process}~\circlednumberblack{4}, Node 14 has two prefix nodes, 6 and 10, each with a distance of 2. 
If both prefixes are selected, two paths would be formed: $2 \rightarrow 6 \rightarrow 14$ and $2 \rightarrow 10 \rightarrow 14$. 
This redundancy leads to unnecessary computations. 
Therefore, we retain only one path by keeping the first prefix.

Subsequently, we set the suffix information and update the prefix node's \textit{Count} to 1 (Lines~8-10). 
For example, Node 6's \textit{Count} is set to 1, designating Node 14 as its suffix. 
If Node 6 has a distance $>1$ and $<4$, the backward pass would continue tracing the path to the first node with distance 1, forming a complete path to Node 14 within a distance of 4.
After that, we will only keep the prefix bitmap with the smallest distance (Line 11).
This mechanism ensures each node maintains a single prefix, facilitating efficient parallelism and preserving correct partial ordering within the Hasse graph.

\textbf{Balanced Forest.}
As shown in \Fig{fig:scoreboard_process}~\circlednumberblack{5}, after the forward and backward passes, we maintain a concise Hasse graph with valid prefixes having the smallest distances, thereby constructing the forest. 
We implement a workload counter and priority supervision to assign each node a Lane ID, ensuring balanced distribution across the trees. For example, Node 15 has two prefixes, Node 7 and Node 14. 
Since Node 2 corresponds to two TransRows, Node 15 is assigned to Lane 1 to achieve load balancing.

\revise{\textbf{Scoreboard Information (SI).}}
\revision{RE3}{Using this approach, we successfully construct a balanced forest through the Scoreboard mechanism.
In practice, the Hasse graph is transformed into a simplified table, referred to as Scoreboard Information (SI), which contains two key elements: each TransRow and its corresponding Prefix, as illustrated in \Fig{fig:scoreboard_process}~\circlednumberblack{6}. }
\revise{
The overhead of SI is minimal. 
The total memory requirement for SI is:  
$
2 \times T \times 2^T \ \text{bits,}
$
where $T$ is the TransRow width.
When $T=8$, the SI needs only 512 Bytes of memory, which is insignificant and stored in the on-chip buffer.
}

\subsection{\revise{Static Scoreboard}}
\label{sec:static}
\revision{CQ2b}{
The execution order for all tensors in a DNN model, including weight and activation tensors, can be precomputed. 
For activation tensors, a small calibration dataset is used to generate the activation tensors.
We first transform the quantized tensor into a binary matrix, which is then divided into $T$-bit TransRows. 
All TransRows within the tensor are recorded, and the final static Scoreboard Information (SI) is computed offline using our proposed algorithm in \Fig{fig:scoreboard_process}.
}

However, static SI may introduce efficiency challenges.  
GEMM relies on tiling (\Sec{sec:tiling}), and for a given tile, the prefix of a TransRow in the static SI may not exist in that tile, leading to an \textbf{SI Miss}, similar to a cache miss.  
An SI miss disrupts the prefix-suffix path, significantly impacting performance.  
We conduct experiments to evaluate this impact in \Sec{sec:static_dynamic}.  
\revise{
To ensure that the Scoreboard remains transparent to users and minimizes SI misses, we propose an efficient hardware-supported \textbf{dynamic Scoreboard} for enhanced performance.  
}

\subsection{\revise{Dynamic Scoreboard}}
\begin{figure}[t]
    \centering
    \includegraphics[width=0.9\linewidth]{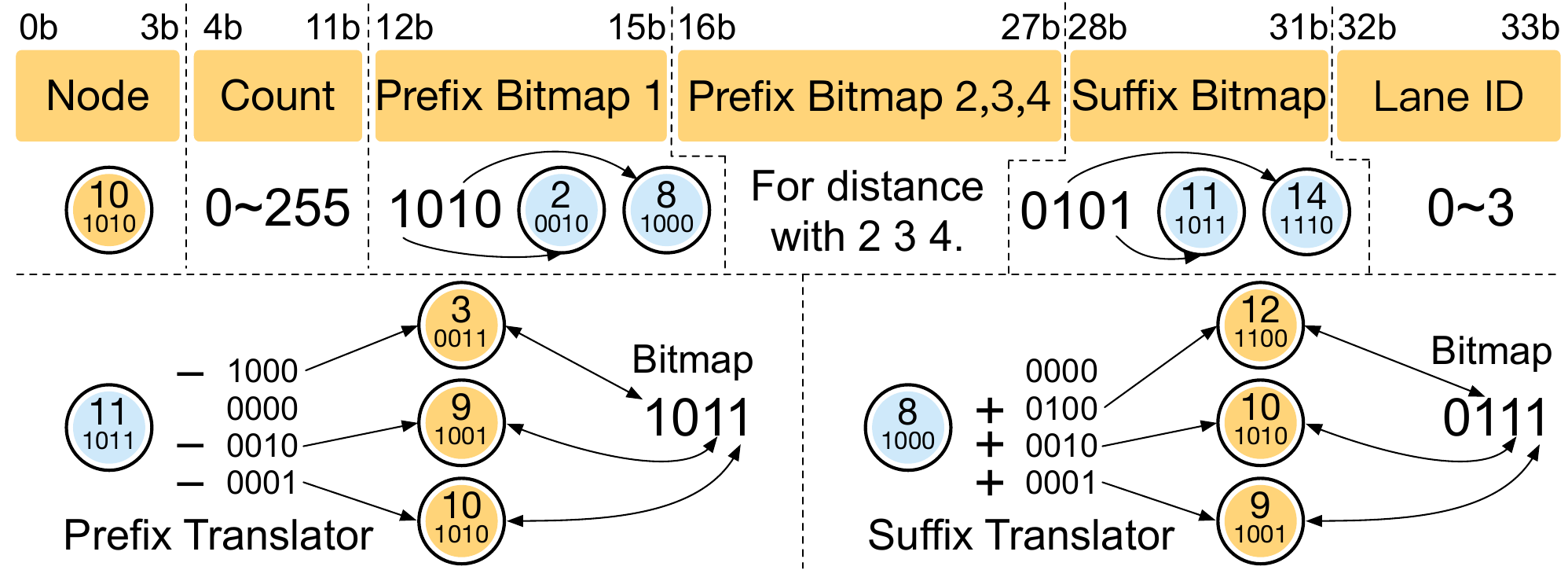}
    \caption{Bit field diagram of 4-bit Scoreboard and prefix and suffix translators.}
    \label{fig:scoreboard_data}
    \vspace{-5mm}
\end{figure}

\Fig{fig:scoreboard_data} illustrates the bit fields of a single entry in the 4-bit \revise{dynamic} Scoreboard design.

\textbf{Node and Count:} Each entry contains a Node identifier (e.g., Node 10) and an 8-bit \textit{Count} field, representing the number of TransRows sharing the same value as the node. The Count field facilitates load balancing across the trees.

\textbf{Prefix Bitmaps (PBs):} We allocate four Prefix Bitmap fields, each corresponding to prefix indices at distances of 1, 2, 3, and 4. 

\textbf{Suffix Bitmap (SB):} The Suffix Bitmap records suffix nodes that are absent from the current set of TransRows. 

\textbf{Lane ID:} Each entry includes a Lane ID field that records the lane identifier associated with the node. This facilitates the distribution of tasks across multiple parallel lanes.

Furthermore, the Scoreboard can be extended to accommodate an 8-bit Hasse graph by maintaining the same fields but increasing the bit-width accordingly to handle the larger number of nodes.

Thanks to the Hasse representation, each node is shown to have a limited set of prefixes and suffixes. 
Leveraging this property, we propose an efficient encoding and decoding mechanism to map bitmap representations into prefix and suffix indices. 
As shown in \Fig{fig:scoreboard_data}-bottom, the proposed method introduces a \textbf{Prefix Translator} and a \textbf{Suffix Translator} to compress and recover transitive sparsity information effectively.
The Prefix Translator derives node indices by applying a 1-to-0 bit flip to the prefix bitmap, while the Suffix Translator reconstructs suffix indices through a 0-to-1 bit flip. 
This design eliminates the need to explicitly store prefix and suffix data, significantly reducing significant ($T$ times for $T$-bit) memory overhead. 
Moreover, the method achieves fast and accurate decoding with minimal hardware complexity, making it highly suitable for high-performance systems.

Finally, we get only one prefix for each node. 
It is important to note that the \textbf{distance} in the algorithm is a conceptual construct. 
In the hardware implementation, a straightforward logic circuit utilizing bitmaps can accurately determine the distance without introducing conflicts or dependencies, thanks to the efficient data structure design.
Consequently, we are able to fully exploit the inherent parallelism of the Hasse graph by processing each level concurrently. 
This is facilitated by a multi-way Scoreboard design, which achieves high efficiency during on-the-fly operations.

\section{Transitive Array}
\revision{CQ1d}{\Fig{fig:arch} and \Fig{fig:arch2} shows Transitive Array design and an example of processing TransArray computation. }

\subsection{Tiling}
\label{sec:tiling}
Tiling is a fundamental technique for partitioning GEMM operations into smaller tiles, optimizing execution for DNN accelerators.
\revision{RA7}{In \Fig{fig:arch2}~\circlednumberblack{1}, we illustrate an example featuring an \revise{$S$-bit weight matrix of shape ($N \times K$)}, an 8-bit input matrix of shape ($K \times M$), and a 32-bit partial sum output matrix of shape ($N \times M$).}
\revise{Initially, tensors are stored in off-chip DRAM.
For each TransArray, only a small tile is processed in the on-chip buffer, consisting of a weight tile ($n \times k$), an input tile ($k \times m$), and an output tile ($n \times m$).
Using the bit-slice method, the weight tensor is decomposed into a binary weight tile of shape ($S \cdot n \times k$).
The pre-quantization and bit-slicing processes are performed offline, while all subsequent steps execute at runtime.
During computation, the TransArray unit processes a sub-tile consisting of a weight tile ($S \cdot n \times T$) and an input tile ($T \times m$), where $T$ represents the TransRow width.
}

\subsection{\revise{Scoreboarding}}
\label{sec:Scoreboarding}
\begin{figure}[t]
    \centering
    \includegraphics[width=0.9\linewidth]{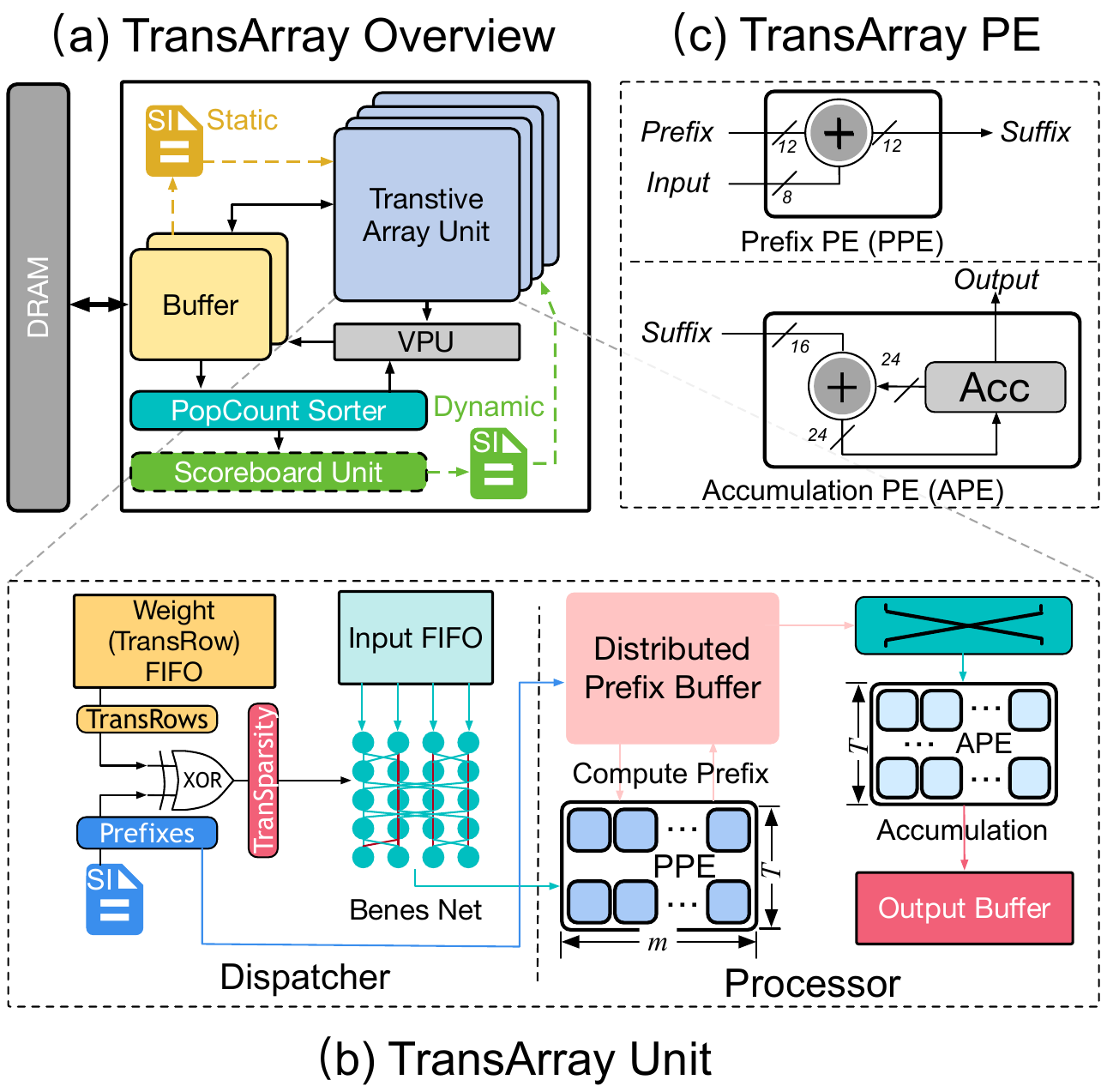}
    \vspace{-3mm}
    \caption{\recaption{CQ1d}{(a) Overview of Transitive Array architecture. (b) One Transitive Array unit. (c) Prefix PE (PPE) and Accumulation PE (APE) design.}}
    \vspace{-6mm}
    \label{fig:arch}
\end{figure}

\revision{CQ2b}{
In \Fig{fig:arch2}~\circlednumberblack{1} bottom, TransArray can select only one type of Scoreboard Information (SI) between static and dynamic.
The \textbf{static SI} is stored in DRAM and pre-fetched by TransArray before GEMM computation, incurring no runtime overhead.  
Thus, \textbf{all sub-tiles in the tensor share the same static SI.}
In contrast, the \textbf{dynamic SI} is generated at runtime using the hardware Scoreboard when the weight sub-tile ($S \cdot n \times T$) is sent to the on-chip network. 
Unlike the static SI, \textbf{each sub-tile has its own private dynamic SI}, enabling optimal prefix for each TransRow.  
As illustrated in In \Fig{fig:arch}~(a), the sub-tile and the Scoreboard unit can be shared by multiple TransArray units.
}

\subsection{Dispatch}
\begin{figure}[t]
    \centering
    \includegraphics[width=1\linewidth]{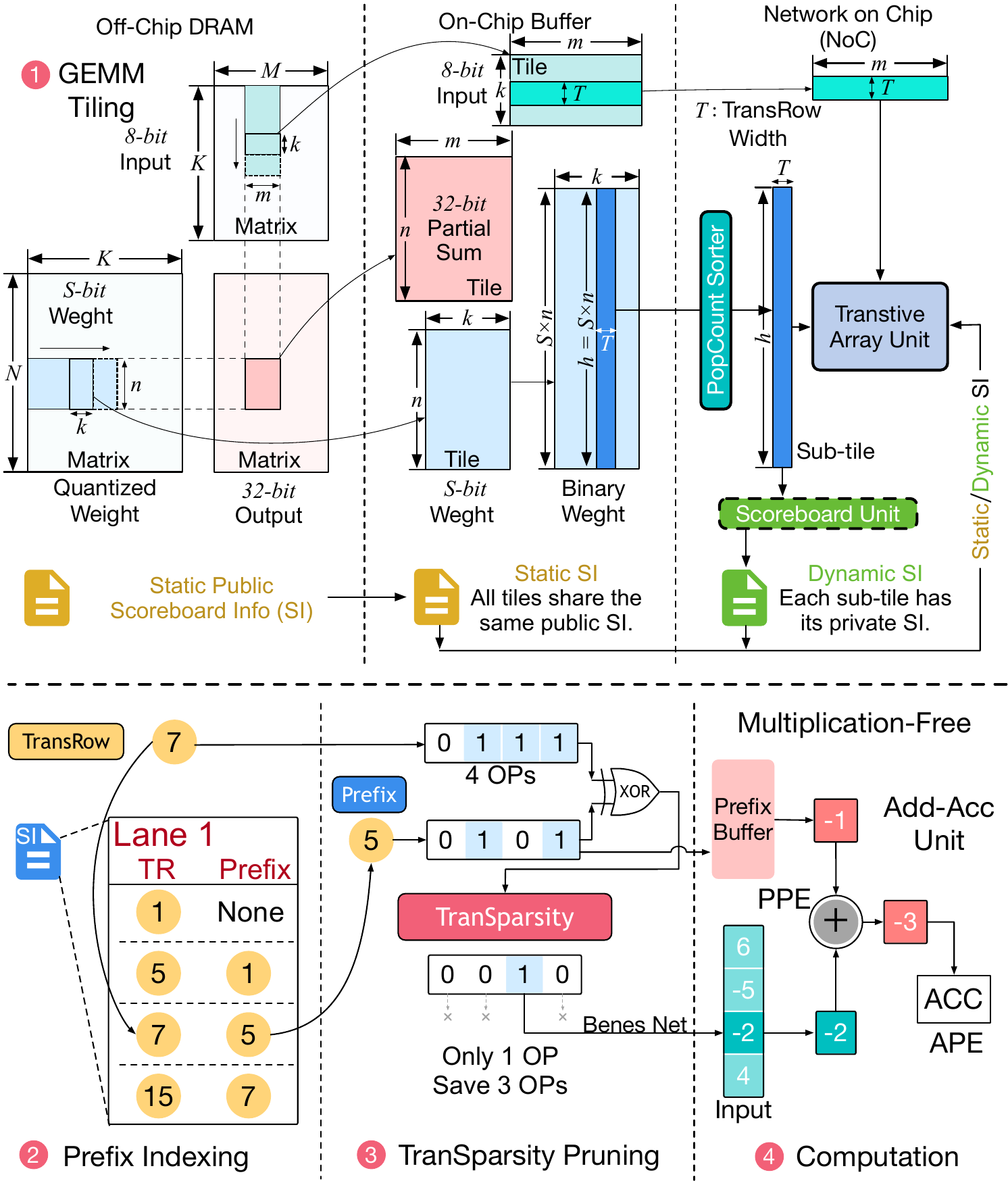}
    \vspace{-6mm}
    \caption{
    \recaption{CQ1d}{
        Processing flow of the Transitive Array unit.  
        \protect\circlednumberblack{1}: GEMM tiling layout and Scoreboard Information (SI).  
        \protect\circlednumberblack{2}: Retrieving the prefix for each TransRow using SI.  
        \protect\circlednumberblack{3}: Pruning for TranSparsity.  
        \protect\circlednumberblack{4}: Computation.
        }
    }
    \vspace{-5mm}
    \label{fig:arch2}
\end{figure}

\revision{CQ1d}{
Next, the TransArray dispatches TransRows along with their corresponding Scoreboard Information (SI) 
in \Fig{fig:arch}~(b) Dispatcher and \Fig{fig:arch2}~\circlednumberblack{2} and \circlednumberblack{3}.
During this dispatch process, {TranSparsity} is computed using a simple XOR gate.  
For example, when the TransArray receives \textbf{TransRow 7} (\texttt{0111}), it first retrieves its \textbf{Prefix 5} (\texttt{0101}) from the SI.  
Then TransArray uses XOR ($\oplus$) to prune TransRow into \textbf{TranSparsity}: $7 \oplus 5 = 2 \ (\texttt{0010})$.  
The TranSparsity value (\texttt{0010}) corresponds to the element in the input matrix, i.e., \textbf{$-2$} in \Fig{fig:arch2}~\circlednumberblack{4}.  
Additionally, the Prefix (\texttt{0101}) is sent to the {Prefix Buffer}, allowing retrieval of the precomputed prefix result, \textbf{$-1$} for Prefix 5 (\texttt{0101}).  
Thus, instead of each TransRow accumulating all 4 input elements as in the original dense GEMM, it now requires only a {single accumulation}, effectively saving {$T\times$ operations for $T$-bit TranSparsity}.  
}

This reduction is the primary source of sparsity in our design. 
As a result, our architecture achieves up to \textbf{75\%} sparsity for 4-bit TranSparsity and \textbf{87.5\%} sparsity for {8-bit configurations}. 
Furthermore, after the first dispatch, each node in the Scoreboard replaces its Prefix with its own index.  
Subsequent identical TransRows can reuse the result from the first computed TransRow.
This significantly improves efficiency.

\subsection{Distribution Network}
\revise{In \Fig{fig:arch} (b) Dispatcher and \Fig{fig:arch2} Step~\circlednumberblack{3}, the TransArray needs to fetch the input data and prefix data.}
We adopt the Benes network~\cite{benes1964mathematical} to support input data access, which is an established non-blocking network widely adopted in studies~\cite{qin2020sigma, waksman1968permutation}.
The Benes network has only $2\log(N) + 1$ levels for an $N$-input and $N$-output configuration, resulting in very high hardware efficiency. 
For the prefix buffer, we employ a distributed buffer design where each prefix buffer operates independently. 
This approach significantly reduces the overhead associated with prefix buffers by eliminating the need to store all prefix nodes and avoiding complex network designs. 

Additionally, we incorporate a crossbar between \revise{Step~\circlednumberblack{3} and Step~\circlednumberblack{4}} for data arrangement in the prefix buffer to avoid bank conflicts originating from the row index. 
For each cycle, only $T$ vectors are sent to the double buffer in $T$-bit TranSparsity.
If the $T$ vectors correspond to the same memory bank based on their row indices, conflicts may arise when they are stored in the same buffer bank. 
To mitigate this, we introduce a queue within the crossbar to buffer the partial sums and arrange them appropriately. 
We implement a double buffer mechanism so that the partial sum buffer overlaps and conceals the overhead associated with buffering, thereby enhancing overall system efficiency.
We also put a shifter in place to shift their value to accommodate their bit-level according to their indices.

\subsection{Processing Element}
\label{sec:pe}
In \revise{Steps~\circlednumberblack{4}}, we utilize adders to complete all computations for the TransArray. 
For the TransArray, we design two types of Processing Elements (PEs).

As shown in \revise{\Fig{fig:arch}~(c)}, the first PE is the \revise{\textbf{Prefix Processing Element (PPE)}}, which is responsible for transitively computing the prefix results. 
According to our analysis in \Sec{sec:prelim}, maintaining sufficiently high precision in the adder allows us to perform lossless computations based on integers. 
Therefore, we employ a 12-bit adder to satisfy the precision requirements for PPE. 
The PPE design is highly concise, featuring only one 12-bit adder that adds the input and prefix sum, stores the result in the prefix buffer.
\revise{The TPE array is formed by $T$ lanes, each with $m=32$ adders for $T$-bit width TranSparsity.}

The second PE is the \textbf{Accumulation Processing Element (APE)}, depicted in \Fig{fig:arch}~\circlednumberblack{9}, which is responsible for accumulating the final results. 
The APE only comprises a \revise{24}-bit accumulator. 
It accesses the partial sums from the prefix buffer and efficiently combines them to produce the final output. 
The APE array has the same shape as the \revise{PPE}, i.e., \revise{$T \times 32$ for $T$-bit} TranSparsity.

\revise{As shown in \revise{\Fig{fig:arch}~(a)}, we incorporate vector units (VPU) to support operations beyond GEMM (e.g., de-quantization, softmax, etc.), similar to previous studies~\cite{guo2022ant,guo2023olive,lee2024tender}.}
According to the latest study~\cite{wan2024efficientlargelanguagemodels}, we utilize group-wise quantization to further improve model accuracy. 
When the group size is $128$, the vector unit applies an integer scale factor to re-scale the partial results for each $128/T$ tile with $T$-bit TranSparsity. 
Therefore, we can efficiently overlap the overhead.

\revision{RA8}{}\revision{RE2}{
TransArray inherently supports the mixed-precision design.
First, for the weight tensor, the bit-slicing method decomposes arbitrary bit-width weight tensors into binary matrices, making mixed-precision design inherently compatible with the TransArray design.  
For the activation tensor, since our processing elements (PEs) only use adders, it is straightforward to configure them for different bit widths.  
To support 8-bit activation, we employ a 12-bit Partial Product Engine (PPE) and a 24-bit Accumulation Processing Engine (APE).  
Additionally, they can be easily split into two 6-bit PPEs and two 12-bit APEs to support 4-bit activation.  
}

\begin{figure*}[t]
    \centering
    \includegraphics[width=1\linewidth]{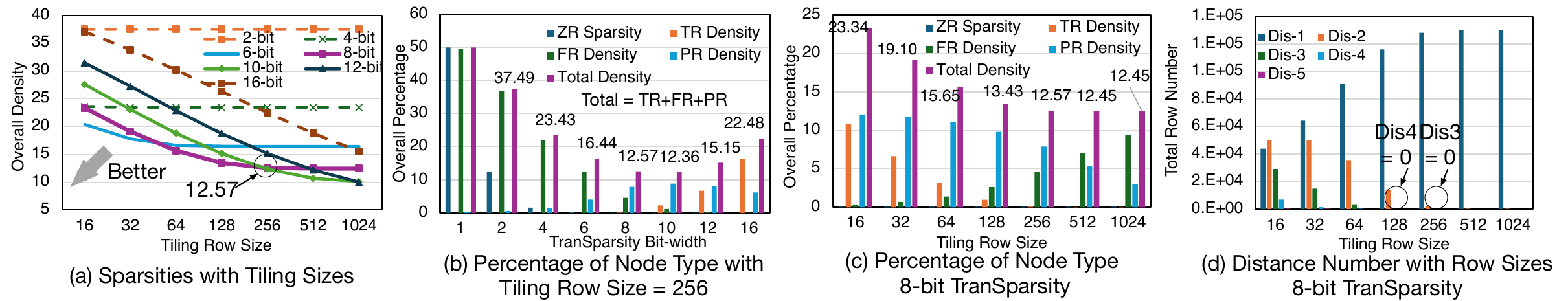}
    \vspace{-6mm}
    \caption{Design space exploration among various bit-width and tiling row sizes: (a) and (b), and 8-bit TranSparsity with tiling row sizes: (c) and (d) on a $1024\times 1024$ random 0-1 matrix.}
    \label{fig:dse}
    \vspace{-2mm}
\end{figure*}

\subsection{Scheduling}
\label{sec:Scheduling}
\revision{RD5}{
Due to the reordering of execution, we implement multiple double-buffering mechanisms to divide the computation into three stages: (1) \revise{Dynamic Scoreboarding}, (2) TPE array for prefix, and (3) APE array for output.}

First, Stages 2 and 3 utilize arrays of identical shapes to compute the prefix and the final output, respectively. 
Specifically, TransRows with a prefix of distance 1 can complete the \revise{PPE} and APE computations within a single cycle. 
However, TransRows with distances $>1$ require additional cycles for the \revise{PPE} to process. 
Consequently, the \revise{PPE} array will consistently require more cycles than the APE. 
For the \revise{PPE}, if we have $n$ TransRows, the number of cycles required will exceed $n$ cycles.
\revise{However, APE will have constantly $n$ cycles.}

For Stage 1, due to the Hamming-weight sort, the dynamic Scoreboard can process a maximum number of nodes given by:
$
\text{min}(n, 2^T)
$
where $n$ is the number of TransRows, and $T$ is the bit-width of TranSparsity. 
To perform the sort, we implement a bitonic sorter~\cite{batcher1968sorting} with $O(\log^2 n)$ time complexity, resulting in significantly fewer cycles. 
Futhermore, we configure an $T$-way Scoreboard to provide parallel Scoreboarding for $T$-bit TranSparsity, ensuring that:
$
\frac{\text{min}(n, 2^T)}{T} < \frac{n}{T},
$
\revise{, which means Scoreboarding time is always less than that of PPE and APE.}

\revise{
Finally, this three-stage pipeline ensures that the critical path of TransArray is still on the PPE array. 
Fortunately, according to our analysis, only approximately $1.67\%$ of TransRows in our design have distances greater than 1. 
Thus, both the \revise{PPE} and APE arrays achieve nearly full utilization.
}

\section{Evaluation}

\subsection{Methodology}

\textbf{Accelerator Baselines.}
We compare our Transitive Array with five baselines, including Bitfusion~\cite{sharma2018bit}, ANT~\cite{guo2022ant}, Olive~\cite{guo2023olive}, Tender~\cite{lee2024tender} and BitVert~\cite{chen2024bbs}. BitFusion~\cite{sharma2018bit} introduces bit-level dynamic composability for DNN acceleration. ANT~\cite{guo2022ant} proposes a hardware-friendly framework for a fixed-length adaptive numerical data type. Olive~\cite{guo2023olive} quantizes models by utilizing the space of normal values to accommodate outliers. Tender~\cite{lee2024tender} decomposes activation tensors along feature dimensions into sub-tensors, with scale factors set to powers of two. BitVert~\cite{chen2024bbs} introduces bi-directional bit-level sparsity, ensuring at least 50\% sparsity, and provides a bit-level binary pruning technique. 

\textbf{Accelerator Implementation.}
We build a cycle-level simulator to analyze the performance of the Transitive Array.
For other architectures, we use their open-sourced simulator, ANT~\cite{guo2022ant}, which all build on. 
The hardware architecture of both the Transitive Array and the baselines is implemented using System Verilog. 
We synthesize the RTL logic with Synopsys Design Compiler, utilizing ARM’s standard cell library in a commercial 28nm process to determine the corresponding area and static/dynamic power consumption. We use Cacti 7.0 to assess buffer area and power at 28nm. We maintained all baselines and the Transitive Array at the same 28nm process and a frequency of 500MHz.
We rewrite all baseline PE implementations according to their design to ensure a fair comparison.
\revise{We evaluate most experiments with dynamic Scoreboard, except the \Sec{sec:static_dynamic}, in which we compare and explore the static and dynamic Scoreboard using random and real data.
}

\textbf{Benchmark.}
We run LLaMA versions 1, 2, and 3~\cite{touvron2023llama1,touvron2023llama2,dubey2024llama3herdmodels} on the Wikitext dataset~\cite{merity2016pointer}, using the perplexity (PPL)~\cite{brown1992estimate} of the generated sentences as the performance metric. Since TransArray's results depend on the data distribution of weights and activations, we extract all the model's weights and activations and evaluate them with our simulator. However, because TransArray requires both activations and weights, resulting in an unacceptable memory footprint trace, we only extract the first Transformer block with a prefill sequence length of 2048. This approach is feasible because all Transformer blocks are identical and exhibit similar computational behavior.

\subsection{Design Space Exploration}
The TranSparsity is a generalized design for arbitrary bit-widths. 
Thus, we first explore the design space to determine our final hardware implementation. 
We employ the tiling approach to divide the large matrix into small tiles. 
Specifically, the performance of the TransArray is significantly impacted by the tile size, i.e., the \revise{row number ($N$) and the bit width ($T$)} of TransRows for a bit matrix.
\revise{All experiments on this sub-section are based on dynamic Scoreboard and random data.
}

\textbf{Bit Width ($T$).}
\revision{RC2}{
As illustrated in \Fig{fig:dse}~(a), we explore the influence of tiling size on $N$ and $T$. 
We find that $T$ determines the upper bound and lower bound of sparsity and density for TranSparsity. 
Even if we fully utilize the sparsity of TranSparsity, we can only achieve a lower-bound density around $\frac{1}{T}$ because we must perform at least one accumulation operation for every $T$-bit element. 
For example, 8-bit TranSparsity requires one accumulation every 8 bits. 
For 8-bit TranSparsity, we achieve \textbf{Pareto Optimality} with reasonable overhead. 
Beyond 8-bit, the density increases before reaching a row size of 256. 
Notably, 10-bit TranSparsity requires $4\times$ the hardware overhead of 8-bit TranSparsity but offers comparable sparsity at a tile row size of 256. 
Therefore, we choose \textbf{8-bit TranSparsity} to implement in this study.
}

As mentioned earlier, not all TransRows will utilize both \revise{PPE} and APE. 
The utilization of \revise{PPE} and APE significantly impacts the overall design efficiency. 
We summarize the features of TransRows based on the utilization of \revise{PPE} and APE for partial sums (PSum) and final sums (FSum). 
This classification helps us understand the efficiency of TranSparsity in depth. 
We identify four computation patterns based on the combination of PSum and FSum.
\textbf{Zero Row (ZR):} Represents a node with no partial or final sums, allowing us to skip computations entirely.
\textbf{Transitive Reuse (TR):} Represents a node responsible for transitioning partial results to its suffix node, utilizing only the \revise{PPE} without APE.
\textbf{Full Result Reuse (FR):} Represents nodes that have been computed previously, allowing subsequent TransRows to reuse the results without requiring prefix computations or \revise{PPE}.
\textbf{Prefix Result Reuse (PR):} Represents nodes that require both \revise{PPE} and APE to add the prefix result and perform accumulation.

As illustrated in \Fig{fig:dse}~(b) with a tiling row size of 256, before 8-bit TranSparsity, the FR (Full Result Reuse) constitutes the majority of TransRows, leading to underutilization of the TransArray and higher density. 
After 8-bit, higher bit widths such as 10, 12, and 16 introduce more TR (Transitive) nodes. 
This results in an extremely sparse Hasse graph consisting of only 256 rows. 
The 8-bit TranSparsity strikes a balanced trade-off between sparsity and hardware overhead.

\begin{table}[b] 
    \vspace{-6pt}
    \caption{Specifications of One TransArray Unit.}
    \vspace{-8pt}
        \centering
    \resizebox{0.42\textwidth}{!}{%
        \begin{tabular}{l|l}
            \toprule
            Bit-width  & $T=8$-bit TranSpasrity.\\ \midrule
            TransRow Number  & Max 256 1-bit TransRow.\\ \midrule
            Weight Tiling & $N=32$ for 8-bit Wgt; $N=64$ for 4-bit Wgt. \\ \midrule
            Input Tiling & $M=32$ for 8-bit Input. \\ \midrule
            \revise{PPE} Array & $8\times 32$ 12-bit Adder. \\ \midrule
            APE Array & $8\times 32$ 24-bit Adder. \\ \midrule
            NoC & An 8-way Benes net and X bar.\\ \midrule
            Scoreboard & Tow 8-way 256-entry tables; A sorter.\\ \midrule
            \multirow{2}{*}{Buffer Size: 80KB} & 8KB Weight; 8KB Input; 22KB Output; \\ 
                         & 18KB Prefix; 24KB Double Buffer. \\
            \bottomrule
        \end{tabular}
        }
        \label{tab:config}
        \vspace{-1mm}
\end{table}

\textbf{Row Number ($N$).}
As shown in \Fig{fig:dse}~(c), we conducted experiments varying the number of rows from 16 to 1024 using 8-bit TranSparsity. 
We observed that when the number of rows exceeds 256, the density of 8-bit TranSparsity stabilizes. 
This stability occurs because most TransRows are captured by the Hasse graph, resulting in only a few Transitive Result (TR) nodes. 
Consequently, increasing the row number beyond 256 does not yield additional sparsity benefits. 
Even if all nodes can reuse results, every 8 bits still requires one addition operation. 
Therefore, we set the \textbf{maximum row number to 256} for 8-bit TranSparsity. 
With this configuration, each TransArray unit can support $32 \times 8$ tile for 8-bit weights and $64 \times 8$ tile for 4-bit weights.

Additionally, we provide detailed statistics on the distances of the nodes. 
As depicted in \Fig{fig:dse}~(c), when the row number is 128, there are no nodes with a Distance of 4, and similarly, for a row number of 256, there are no nodes with a Distance of 3. 
In practice, this means we only need to provide three prefix bitmap fields for nodes with a Distance of 3. 
TransRows with a Distance $\geq 4$ are treated as outliers and dispatched at the end of other operations.

\subsection{Area Comparison}
Table~\ref{tab:config} and Table~\ref{tab:area} present the design configurations and area analysis of our TransArray architecture compared to the baseline designs. 
We implement \textbf{\revise{six} TransArray units} in the final GEMM accelerator design.
TransArray yields a lower computational core area overhead compared to all five baselines. This low area overhead mainly comes from our simplified \revise{PPE} and APE design, which only requires accumulation operations, while other baselines employ multi-bit multiplication units, incurring quadratic hardware complexity. Therefore, we are able to place a NoC within our array to flexibly support buffer access while maintaining a lower area overhead.
\revision{CQ3}{Note that we also consider the Scoreboard to ensure a dynamic Scoreboard.} 
To ensure a fair comparison, we configure a smaller buffer size (480KB) for our design compared to other baselines (512~KB and 608~KB).

\begin{table}[b] 
    \vspace{-6pt}
      \caption{{\recaption{CQ3}{The area of core components and buffers for TransArray and other baseline models using a 28nm process.}}}
      \vspace{-4pt}
    
      \renewcommand{\arraystretch}{1.2}
      \resizebox{0.48\textwidth}{!}{%
      \begin{tabular}{c|lcc|c}
      \toprule
      \multirow{2}{*}{Arch.} & \multicolumn{3}{c|}{Computation Core}  & \multirow{2}{*}{Buffer}  \\ 
    
      & Component & Array & Area ($mm^2$) &  \\ 
      \toprule
      
      & \revise{PPE} (50.3$\mu m^2$) & $\revise{6}\times(8\times 32)$ & \multirow{4}{*}{\revise{\textbf{0.443}}} &\multirow{4}{*}{$480$KB}  \\ 
      \multirow{1}{*}{TransArray}& APE (101.7$\mu m^2$) & $\revise{6}\times(8\times 32)$ &  &  \\ 
      (\revise{6} Units)& NoC (19520$\mu m^2$) & $\revise{6}\times (1)$ &  & \\
      & \revise{Socreboard (92507$\mu m^2$)} & \revise{$1$} &  &  \\ \cline{1-5}
      
      BitFusion & 8-bit PE (548$\mu m^2$) & $28\times 32$ & 0.491 &\multirow{4}{*}{$512$KB}   \\ 
  
      ANT& 4-bit PE (210$\mu m^2$) & $36\times 64$ &{0.484} &  \\ 
      
      Olive & 4-bit PE ({319}$\mu m^2$) & $32\times 48$ &{{0.489}}  &  \\

      BitVert & 8-bit PE (985$\mu m^2$) & $16\times 30$ & 0.473  &   \\ 
      \cline{2-5}
      Tender & 4-bit PE (329$\mu m^2$) & $30\times 48$ & 0.474  & \revise{608KB} \\

    \bottomrule
      \end{tabular}%
      }
      \vspace*{-1mm}
      \label{tab:area}
  \end{table}

\begin{figure*}
    \centering
    \includegraphics[width=0.99\linewidth]{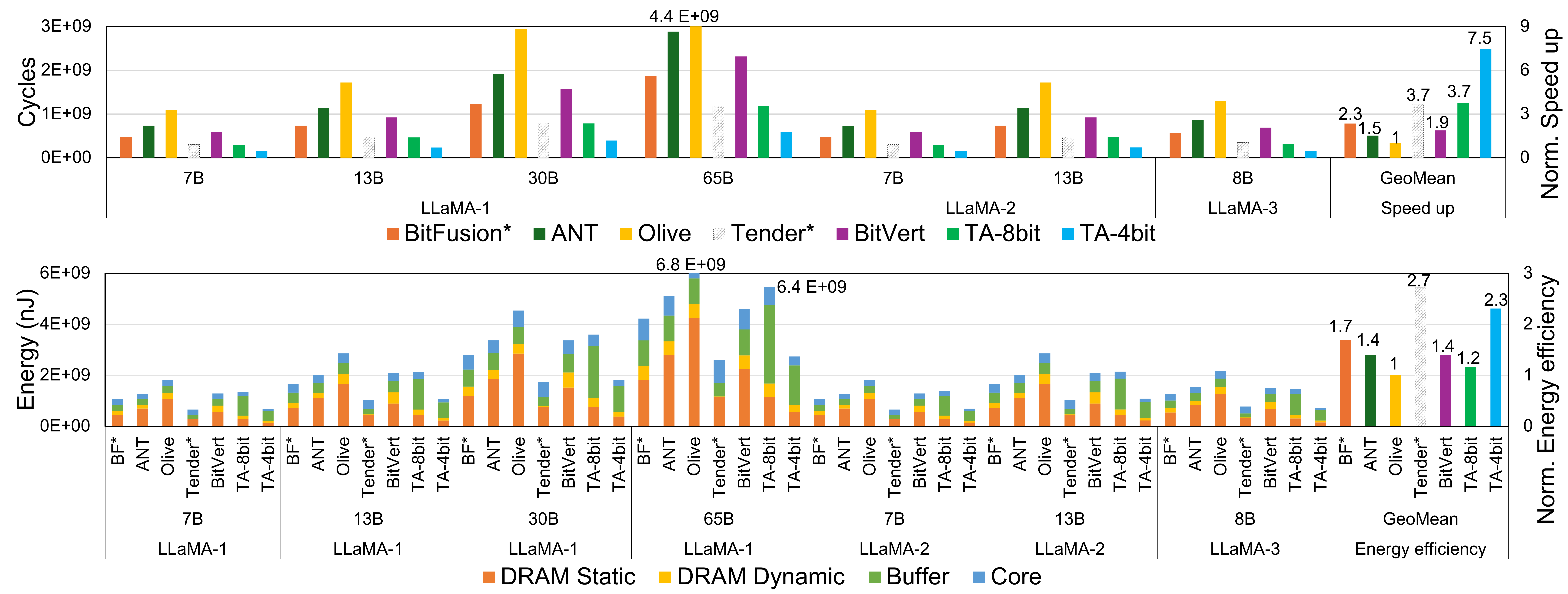}
    \caption{\recaption{CQ3}{} Runtime and energy consumption on \recaption{CQ4a}{FC layers} of LLaMA models. *BitFusion (8-bit) and Tender (4-bit) exhibit unacceptable perplexity (PPL) scores. ANT, Olive, BitVert, and TransArray (4-bit) achieve similar PPL levels on LLaMA.}
    \label{fig:main}
    \vspace{-1mm}
\end{figure*}

\begin{table}[t]
  \centering
  \small
  \renewcommand{\arraystretch}{1.2}
  \caption[]{\protect\revision{RA6}{} The Perplexity on Wikitext data with Tender (TD), BitFusion (BF), Olive (OL), BitVer (BV), ANT with group quantization, and TransArray (TA) with group quantization.
  }
  \vspace*{-0.3cm}
  \resizebox{1\columnwidth}{!}{
  \scriptsize
    \begin{tabular}{l|c|ccccccc|c}
      \toprule
      \textbf{Arch} & {TD-4} & {BF} & {OL} & {TD-8} & {BV} & {ANT} & \textbf{TA} & \textbf{\revise{TA}} & FP16 \\
      \midrule
      \textbf{FC Act.} & 4bit & 8bit & 8bit & 8bit & 8bit & 8bit & \textbf{Int8} & \revise{Int8} & FP16 \\
      \textbf{FC Wgt.} & 4bit & 8bit & 8bit & 8bit & 8bit & 8bit & \textbf{Int4} & \revise{Int8} & FP16 \\ \hline
      \textbf{L-1 7B} & 23.85 & 9.50 & 5.86 & 5.87 & --- & \textbf{5.82} & \textbf{5.82} & \revise{5.75} & 5.68\\
      \textbf{L-1 13B} & 13.68 & 8.46 & 5.28 & 5.28 & --- & \textbf{5.20} & \textbf{5.20} & \revise{5.14} & 5.09\\
      \textbf{L-1 30B} & 12.07 & 6.70 & 4.37 & 4.27 & --- & 4.32 & \textbf{4.24} & \revise{4.17} & 4.10\\
      \textbf{L-1 65B} & 8.85 & 5.34 & 3.80 & 3.74 & --- & 3.76 & \textbf{3.66} & \revise{3.57} & 3.53 \\\hline
      \textbf{L-2 7B} & 36.47 & 10.68 & 5.73 & 5.77 & --- & \textbf{5.58} & 5.62 & \revise{5.56} & 5.47\\
      \textbf{L-2 13B} & 55.08 & 16.11 & 5.06 & 5.09 & --- & 5.20 & \textbf{5.01} & \revise{4.95} & 4.88 \\ \hline
      \textbf{L-3 8B} & 28.60 & 22.56 & 6.70 & 7.17 & \textbf{6.24} & 6.27 & 6.59 & \revise{6.39} & 6.14\\
      \bottomrule
    \end{tabular}
    }
    \label{tbl:ptq_result_transposed}
    \vspace*{-0.6cm}
  \end{table}

\subsection{Model Accuracy}
We benchmarked several post-training quantization (PTQ) methods on the Wikitext dataset using LLaMa models, as shown in \Tbl{tbl:ptq_result_transposed}. 
Perplexity (PPL) served as the performance metric—lower values indicate better performance. 
We reproduced their results using open-sourced code, except for BitVert~\cite{chen2024bbsbidirectionalbitlevelsparsity}, for which we reported only the available results from its paper. 
Due to the lack of optimization for quantization, BitFusion exhibits a larger gap compared to the FP16 results. 
Other accelerators achieve near-lossless performance at the 8-bit level due to their quantization-aware optimized architecture designs. 
We modified ANT to support group-wise quantization for a fair comparison. 
Please note that Tender~\cite{lee2024tender} only supports 4-bit PEs without mixed precision. 
However, its 4-bit PPL is unacceptable. 
Therefore, in the following evaluation, only the results of BitFusion and Tender are provided for reference. 
Additionally, please note that these accelerators (except BitFusion) cannot support the Attention layer due to their complicated weight pre-processing. 
We maintain the Attention layer with FP16 precision.

Due to the generalized design of \name, it can broadly support state-of-the-art (SOTA) quantization frameworks without specific requirements. 
We implement TransArray in Qserve~\cite{lin2024qservew4a8kv4quantizationcodesign}, the SOTA quantization framework from prior work. 
In contrast, other frameworks present a high barrier to algorithmic optimization due to their specialized designs for specific data types. 
\revision{RA9}{ 
For instance, 
SOTA quantization methods, such as SmoothQuant~\cite{xiao2024smoothquantaccurateefficientposttraining}, suppress outliers in weight tensors for integer optimization.  
This contrasts with Olive, which benefits from large outliers.
}

Our proposed \name\ architecture consistently achieves competitive PPL scores across all model sizes, utilizing {Int4 weights and Int8 inputs}, and without Attention quantization for a fair comparison. 
These significant improvements stem from the generalized integer-based design without special requirements.

\subsection{Performance and Energy on \revise{FC Layers}}
\textbf{Settings.}
\Fig{fig:main} shows the overall performance and energy consumption across various accelerators \revision{CQ4a}{\textbf{on only FC layers of the LLaMA models}}. 
The results for BitFusion with 8-bit PE and Tender with 4-bit PE are provided only for reference due to their large PPL loss. 
The remaining accelerators maintain similar accuracy levels for LLaMA. 
ANT and Olive utilize mixed-precision designs with 8-bit evaluations. 
BitVert employs bit-slice techniques with 50\% sparsity. 
We provide two types of precision for TransArray: 4-bit weights and 8-bit weights \recaption{RA6}{on real data}. 

\textbf{Iso-Precision Comparison.}
We first compare TransArray with 8-bit weights against other baselines. 
Due to the greater difficulties in quantizing LLM, the mixed-precision advantages of ANT and Olive disappear. 
They are even slower than BitFusion because of the larger overhead from their complex PEs, which support outliers and adaptive data types. 
Although BitVert has a more complicated PE design, it utilizes bit sparsity with a fixed 50\% sparsity based on bit-slice techniques and achieves a 1.9$\times$ speedup over Olive on LLMs, consistent with the results reported in its paper. 
Despite TransArray needing to slice 8-bit into 8 accumulation operations, it can fully exploit 87.5\% sparsity with TranSparsity. 
This translates to $8\times$ and $4\times$ speedups over dense GEMM and bit sparsity-based architectures, respectively. 
Additionally, with an extremely streamlined PE design, TransArray with 8-bit weights achieves \revise{$2.47\times$}, \revise{$3.75\times$}, and \revise{$1.99\times$} speedups over ANT, Olive, and BitVert, respectively, while maintaining similar energy consumption.

\textbf{Iso-Accuracy Comparison.}
Due to the generality of TransArray, it can employ state-of-the-art (SOTA) algorithms to further enhance its performance. 
For 4-bit quantization, TransArray can theoretically provide $16\times$ and $8\times$ improvements over 8-bit quantization-based and bit-sparsity-based architectures, respectively. 
TransArray with 4-bit weights achieves \revise{$4.91\times$}, \revise{$7.46\times$}, and \revise{$3.97\times$} speedups and \revise{$1.65\times$}, \revise{$2.31\times$}, and \revise{$1.65\times$} energy efficiency improvements over ANT, Olive, and BitVert, respectively.
These significant improvements are also attributed to the generalized design of TransArray. 
Other architectures find it difficult to benefit from algorithmic advancements.

\begin{figure}
    \centering
    \includegraphics[width=0.85\linewidth]{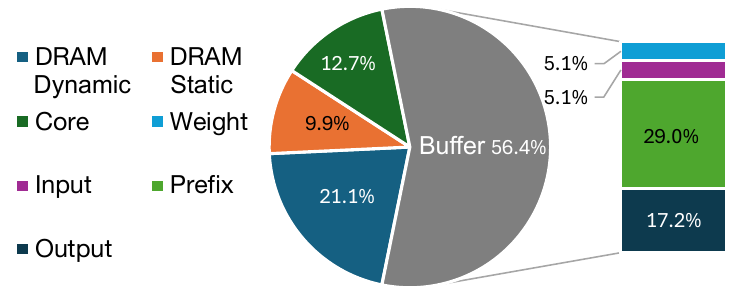}
    \vspace{-3mm}
    \caption{\recaption{CQ3}{TransArray energy breakdown on LLaMA-1-7B.}}
    \vspace{-3mm}
    \label{fig:breakdown}
\end{figure}

\subsection{\revise{Energy Breakdown Analysis}}
\label{sec:energy}
\revision{RB2}{
\Fig{fig:breakdown} presents the energy breakdown of TransArray for the first fully connected (FC) layer of the LLaMA-1-7B model.  
We observe that buffer operations consume the majority of the energy, primarily due to frequent accesses to the {prefix buffer}, which is essential for enabling efficient TranSparsity. 
Although our design incurs significant on-chip buffer access overhead, the high efficiency of TranSparsity significantly reduces overall execution time.  
As a result, TransArray achieves lower {DRAM static energy consumption} compared to other baselines.  
Consequently, despite increased buffer access, TransArray maintains high energy efficiency relative to these baselines. 
}

Essentially, the TransArray design enhances computational efficiency at the expense of increased buffer energy consumption. 
Furthermore, TransArray resembles SIMD~\cite{Flynn1972SomeCO} or in-memory processing~\cite{chi2016prime, jeon2017hmc, hsieh2016transparent} architectures. We may implement TransArray using GPU tensor cores~\cite{nvidia2020a100} to explore broader scenarios and leverage advanced technologies for more energy-efficient solutions.

\subsection{\revise{Performance on Attention Layers}}  
\label{sec:attention_perf}  

\revision{CQ4c}{} \revision{RD6}{  
\Fig{fig:attention} presents the performance results of the Attention layers in LLaMA-1-7B, LLaMA-2-2B, and LLaMA-3-8B.  
All data are collected from real model inference with a sequence length of 2048.  
For Attention layers, we treat the K and V cache as weight tensors.  
Since Attention is challenging to quantize, we adopt {8-bit group-wise quantization} for both ANT and TransArray, while using {16-bit quantization} for BitFusion. 
TransArray achieves a $1.54\times$ speedup over ANT under the same quantization settings and a $3.97\times$ speedup over BitFusion.  
}  

\revise{Furthermore, prior designs~\cite{chen2024bbs, guo2023olive, lee2024tender}—such as Olive, Tender, and BitVert—do not support Attention layers but only focus on fully connected (FC) layers.  
These approaches rely on specific offline quantization methods, making them incompatible with dynamic Attention processing.  
For instance, BitVert depends on complex bit-level binary pruning and channel reordering techniques, which do not support online processing.  
In contrast, TransArray with the dynamic Scoreboard is a general method that can be applied to various integer quantization techniques.  
With the help of the Scoreboard unit, TransArray seamlessly processes all GEMM operations in DNNs while remaining transparent to users.}

\subsection{\revise{Static and Dynamic Scoreboard Comparison}}  
\label{sec:static_dynamic}  
\begin{figure}[t]
    \centering
    \begin{minipage}{0.48\linewidth}
        \centering
        \includegraphics[width=\linewidth]{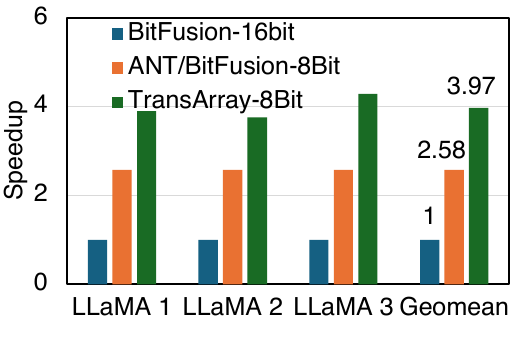}
        \caption{\revise{Speedups on Attention layers of LLaMA models over BitFusion.}}
        \vspace{-5mm}
        \label{fig:attention}
    \end{minipage}
    \hfill
    \begin{minipage}{0.48\linewidth}
        \centering
        \includegraphics[width=\linewidth]{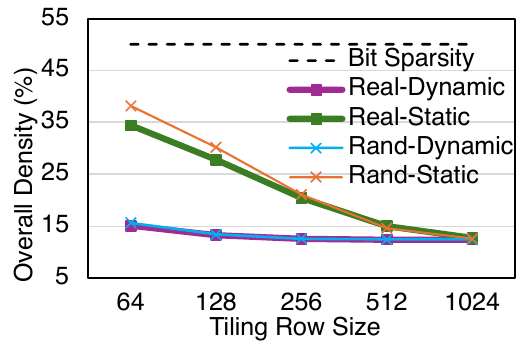}
        \caption{\revise{Static and dynamic Scoreboard comparison on real and random (Rand) data}.}
        \vspace{-5mm}
        \label{fig:static_dynamic}
    \end{minipage}
\end{figure}

\revision{CQ2c}{In this subsection, we compare the static and dynamic Scoreboards on the first FC layer of LLaMA-1-7B model using 8-bit TranSparsity, as illustrated in \Fig{fig:static_dynamic}.}
\revise{Clearly, the dynamic Scoreboard achieves significantly lower density (higher speedup) than the static Scoreboard for smaller tiling row sizes ($<512$), while achieving comparable results for larger sizes ($>512$).}  

 \revise{
A static Scoreboard Information (SI) is shared across an entire tensor.  
When the tile row size is small (e.g., 64), the Hasse graph becomes significantly sparse under 8-bit TranSparsity, containing 256 nodes, of which fewer than 25\% are present.  
This means that different tiles are likely to contain distinct nodes, generating entirely different forests (multiple trees) for SI.  
As a result, static SI leads to frequent \textbf{SI Misses}, causing substantial performance degradation.  
Despite this limitation, the static Scoreboard remains significantly more efficient than bit sparsity. 
However, as the row size increases beyond 256, the static Scoreboard achieves comparable performance to the dynamic Scoreboard, reaching equivalent performance at a row size of 1024.  
Additionally, since the static Scoreboard does not require a dedicated hardware Scoreboard unit, it reduces area overhead by approximately 25\%, leading to potentially better overall performance than the dynamic Scoreboard in some cases.  }

\revise{
In contrast, the dynamic Scoreboard generates SI for each sub-tile, allowing it to achieve near-optimal sparsity and demonstrating strong applicability.  
More importantly, the dynamic Scoreboard is transparent and seamlessly compatible with existing frameworks.  
}

\subsection{\revise{Random and Real Data Comparison} }
\label{sec:random_real}  
\revision{RA6}{
\Fig{fig:static_dynamic} also compares the impact of data distribution on performance.  
We collect real data from the LLaMA-1-7B model and generate 0-1 uniform random data for both static and dynamic Scoreboards.  
Interestingly, TransArray achieves slightly better performance on real data compared to random data.  
After a thorough examination and detailed analysis, we identify two key reasons for this performance improvement:  
(1) Although most original tensors of real data follow a Gaussian distribution, the binary 0-1 distribution after bit-slicing resembles a uniform distribution.  
(2) Certain patterns may exist in the weight tensor of DNN models.  
For uniform random data, the occurrence of repeated values follows the classical problem of discrete uniform distributions. The expected number of unique values in 256 random 8-bit TransRows is approximately 162.  
However, in real data, this number is slightly lower than 162, suggesting the presence of structural patterns in DNN weight tensors.  
}

\subsection{\revise{TransArray in DNN Models} } 
\label{sec:resnet18}  
\begin{figure}[t]
    \centering
    \includegraphics[width=1\linewidth]{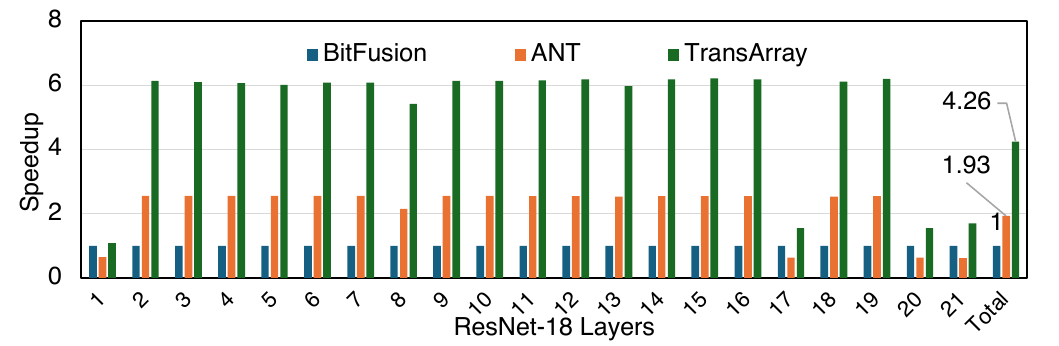}
    \vspace{-6mm}
    \caption{\revise{Speedup comparison on ResNet18 models.}}
    \label{fig:resnet}
    \vspace{-5mm}
\end{figure}

\revision{RB1}{
We evaluate TransArray on the ResNet-18~\cite{he2016deep} model using the ImageNet dataset~\cite{deng2009imagenet}.  
Both BitFusion and ANT are optimized for CNN models with mixed precision.  
For TransArray, we adopt 4-bit quantization using MQBench~\cite{li2021mqbench}.  
TransArray supports mixed-precision computation for 4-bit and 8-bit activations, as discussed in \Sec{sec:pe}.  
We configure the first convolution layer and the final fully connected (FC) layer with 8-bit quantization.  
Following prior work~\cite{guo2022ant}, we also employ \texttt{im2col}~\cite{chetlur2014cudnnefficientprimitivesdeep} to transform convolution layers into GEMM operations.  
\Fig{fig:resnet} presents the speedup results of all ResNet-18 layers for BitFusion, ANT, and TransArray.  
TransArray demonstrates superior performance, achieving a $4.26\times$ speedup over BitFusion and a $2.21\times$ speedup over ANT.  
}

\section{Related Works}
\subsection{DNNs and Quantization}
Deep Neural Networks (DNNs) and Large Language Models (LLMs) have achieved state-of-the-art performance across various domains. 
However, their high computational and memory demands pose challenges in resource-constrained environments. 
Quantization techniques~\cite{yao2020zeroquant, guo2022squant, xiao2024smoothquantaccurateefficientposttraining, guo2025survey, lin2024awqactivationawareweightquantization, ashkboos2024quarotoutlierfree4bitinference, lin2024qservew4a8kv4quantizationcodesign, liu2024spinquantllmquantizationlearned, shao2024omniquantomnidirectionallycalibratedquantization} mitigate these issues by reducing the precision of weights and activations, enhancing computational and memory efficiency with minimal performance loss. 
Notable methods include SmoothQuant~\cite{xiao2024smoothquantaccurateefficientposttraining}, which introduces per-channel scaling; AWQ~\cite{lin2024awqactivationawareweightquantization}, which refines this with activation-aware scaling. These techniques collectively address key quantization challenges in LLMs.

Quantization-based accelerators~\cite{sharma2018bit, zadeh2020gobo, guo2022ant, hu2025m, guo2023olive, liu2025vq} leverage reduced bit-width data representations to minimize memory bandwidth and enhance computational efficiency, making them suitable for resource-constrained environments. 
BitFusion~\cite{sharma2018bit} introduces a flexible PE array that supports various bit-widths dynamically. OLAccel~\cite{park2018energy} employs a heterogeneous quantization strategy with 16-bit MAC units for the first layer and 4-bit MAC units for subsequent layers, optimizing energy efficiency without significant accuracy loss. 
GOBO~\cite{zadeh2020gobo} focuses on outlier-aware quantization by quantizing outlier weights with higher precision. 
ANT~\cite{guo2022ant} proposes a fixed-length adaptive quantization framework but overlooks outliers, limiting its effectiveness. 
OliVe~\cite{guo2023olive} introduces an outlier-victim pair quantization technique that efficiently handles outliers with minimal hardware overhead, achieving negligible accuracy loss with 4-bit quantization. However, this datatype-based quantization accelerator struggles with LLM models, highlighting the need for improved quantization approaches.

\subsection{Bit-slicing and Sparsity}
Therefore, recent techniques~\cite{pragmatic, bitlet, bitwave} have focused on exploiting bit-level sparsity to enhance quantization efficiency. Bit-slice accelerators specifically optimize computations in sparse-bit scenarios by skipping ineffectual zero bits, thus significantly improving efficiency. Pragmatic~\cite{pragmatic} pioneered this approach by selectively processing non-zero bits using variable shifters to align bit significances, simplifying computations but introducing higher hardware overhead. However, such methods are inherently limited to exploiting intrinsic sparsity, typically constrained to approximately 50–60\% due to inherent data characteristics~\cite{chen2024bbsbidirectionalbitlevelsparsity}.

Consequently, many accelerators have turned their attention to sparsity-aware processing~\cite{han2015deep, wen2016learning, guo2020accelerating, zhang_h_2o_2023,  guo2024accelerating, wang2021dual, zhang2024dstc, han2016eie, zhu2019sparse, wei2025prosperity}, though often encountering significant irregularities in memory access patterns.  
To address these irregularities, substantial efforts have been made to regularize sparsity patterns into hardware-friendly formats, such as sparse tensor cores~\cite{zhu2019sparse}, thereby achieving greater speedups on sparse DNN models. BitVert~\cite{chen2024bbsbidirectionalbitlevelsparsity}, a recent bit-slice and sparsity co-design accelerator, dynamically balances workload distribution by selectively skipping zero bits in sparse bit-columns, at the cost of more intricate PE architectures.
Prosperity~\cite{wei2025prosperity} can directly support the bit-level sparsity on spiking neural networks.
In our study, we leverage insights from structured sparsity processing to regularize transitive sparsity, resulting in higher hardware utilization and efficiency gains in our proposed design.
\section{Conclusion}
In this study, we introduced a novel sparsity paradigm, \textbf{Transitive Sparsity}, which leverages the reuse of previously computed results to significantly reduce GEMM computations, thereby decreasing operational overhead. To effectively exploit this sparsity, we designed the \textbf{Transitive Array}, a multiplication-free accelerator that addresses challenges related to execution order and parallelism while also supporting on-the-fly quantization of Attention layers. Comprehensive evaluations demonstrated that the Transitive Array achieves significant speedup compared to state-of-the-art accelerators, maintaining comparable model accuracy on LLaMA models. These improvements are attributed to the generalized design of TransArray, which allows broad support for state-of-the-art quantization frameworks without specialized requirements, and its streamlined PE design that eliminates traditional multiplication operations. 
Our work provides a novel perspective on sparsity optimization in GEMM computation.

\section*{Acknowledgements}
This work was supported in part by NSF-2112562 and ARO W911NF-23-2-0224. The authors sincerely thank the anonymous reviewers for their constructive feedback and valuable suggestions that greatly improved the quality of this work.

\bibliographystyle{ACM-Reference-Format}
\bibliography{refs}

\end{document}